\documentclass[aps,pra,twocolumn,groupedaddress]{revtex4}
\usepackage{graphicx}
\usepackage{amsmath}
\usepackage{color}

\begin{document}

\title{Toward scalable information processing with ultracold polar molecules in an electric field: a numerical investigation} 

\author{La\"etitia Bomble$^1$, Philippe Pellegrini$^1$, Pierre Ghesqui\`ere$^1$ and Mich\`ele Desouter-Lecomte$^{1,2}$}

\affiliation{1 Laboratoire de Chimie Physique, Universit\'e Paris-Sud, UMR 8000, Orsay F-91405, France}
\affiliation{2 D\'epartement de Chimie, Universit\'e de Li\`ege, B\^at B6c Sart Tilman, 400 Li\`ege Belgium}

\date{\today}

\begin{abstract}

We numerically investigate the possibilities of driving quantum algorithms with laser pulses in a register of ultracold NaCs polar molecules in a static electric field. 
We focuse on the possibilities of performing scalable logical operations by considering circuits that involve intermolecular gates (implemented on adjacent interacting molecules) to enable the transfer of information from one molecule to another during conditional laser-driven population inversions.
We study the implementation of an arithmetic operation (the addition of 0 or 1 on a binary digit and a carry in) which requires population inversions only and the Deutsch-Josza
algorithm which requires a control of the phases.
Under typical experimental conditions, our simulations show that high fidelity logical operations involving several qubits can be performed in a time scale of a few hundred of
microseconds, opening promising perspectives for the manipulation of a large number of qubits in these systems. 

\end{abstract}


\maketitle

\section{Introduction}

The past few years have witnessed remarkable experimental achievements on the realization of elementary quantum logical operations on physical systems \cite{Isenhower2010,Monz2009,DiCarlo2009,Politi2009,Du2010}. 
Building a quantum processor that could handle a large number of quantum bits (qubits) represents the next milestone to reach toward the realization of a practical quantum
computer, but no technology is currently available for processing about one hundred of qubits which is the expected minimum number of qubits required to overcome powerfull current
classical computers. Due to their rich inner energy structure that can be used to encode information,  molecules offer promising propects for scalable quantum information processing and 
have attracted a lot of attention recently.
Following the pionnering work of de Vivie-Riedle and
coworkers \cite{Tesch2001,Tesch2002}, several groups have explored the possibily of encoding qubits in ro-vibrational states of a single diatomic molecule 
\cite{Vala2002,Babikov2004,Ohtsuki2005,Menzel2007,Shioya2007, Mishima2008, Tsubouchi2008,Tsubouchi2008b, Sugny2009, Mishima2010, Zaari2010} or polyatomic molecule\cite{Tesch2004,Troppmann2005,Korff2005,Sugny2006,Sugny2007,Ndong2007,Weidinger2007,Bomble2008b,Bomble2009,Bomble2010,Schroder2009} 
or in two interacting diatomic molecules \cite{Mishima2009,
Mishima2009b}. Logic gates were driven by pulses designed by optimal control or genetic algorithms or by using Stimulated Raman Adiabatic Passage (STIRAP) techniques. 
From the viewpoint of scalability, it is challenging to increase the number of encoded qubits on a single
molecule because the number of eigenstates individually addressable cannot grow exponentially. A more
promising strategy would be to use a network of polar molecules holding each a limited number of qubits \cite{DeMille2002}.
Polar molecules are ideal systems for such strategy for they can interact through the strong anisotropic long range
dipole-dipole interaction enabling couplings between qubits to create entanglement \cite{DeMille2002,Carr2009}. Furthermore, experimental progresses obtained in the formation of stable heteronuclear alkaline molecules at ultracold temperatures ($\sim \mu$K) open up the possibilies for an individual 
tight confinement and manipulation of the molecules by optical means paving the way to the actual realization of a polar molecule based quantum register \cite{DeMille2002}.
Different schemes have been proposed to realize universal quantum gates and manipulate qubits encoded on molecular levels of ultracold polar molecules. 
The realization of a conditional phase gate has been proposed by using either a switchable dipole-dipole interaction \cite{Yelin2006,Kuznetsova2008} or a sequence of laser pulses \cite{Charron2007}.

In this article, we present a numerical investigation of the possibilities of driving quantum algorithms with laser pulses on a register of polar molecules trapped in
an optical lattice and experiencing a static electric field. A step toward scalability is addressed by considering schemes that involve intramolecular (implemented on a single molecule) and
intermolecular (when interacting molecules are used) quantum gates. Encodings that use both the vibrational and rotational modes of the molecules are considered as well. We first
implement the addition of 0 or 1 to a binary digit $b_i$ and a carry in $c_i$ to obtain the sum $s_i$ and the carry out $c_{i+1}$\cite{Beckman1996}. This adder requires three qubits and never
involves superposed states. In this sense the computation is weakly affected by dephasing. By omitting the phase constraint, the conditional population inversions can be realized by $\pi$-pulses. Implementing these arithmetic operations on molecular systems at this quasi-classical level should already be appealing for future 
applications. This logical scheme allows for the concatenation of several arithmetic operations by saving the sum $s_i$ in the vibration of one molecule for a latter reading and most importantly, by transfering the carry out $c_{i+1}$ to
a neighboring molecule enabling the next cycle of addition to be carried out without any intermediary readings. A carefull treatment of the
phases for quantum gate involving superposed states is realized in the context of the Deutsch-Josza algorithm \cite{Deutsch1992} applied on a one-bit function.
Similarly to the intermolecular gates of the adder, the active qubit and the ancillary one are
encoded in two neighbouring molecules (see also \cite{Mishima2009} for a two-molecule implementation). Pulses are then designed by optimal control theory (OCT) with an additional constraint for the phase \cite{Tesch2001,Tesch2002}.

This article is organized as follows: first we give a detailed description of the polar molecule based quantum register
\cite{DeMille2002} we consider as a support for realistic numerical simulations and the theoretical framework of our treatment. Then we present the conditional
population inversion driven by $\pi$ pulses and the realization of a three-qubit 0- and 1-adder. Finally, we present the simulation for phase correct gates with application to the Deutsch-Josza algorithm.

\section{Model}
\label{model}
\subsection{Quantum register of polar molecules}

\begin{figure}
\includegraphics[width=0.9\columnwidth]{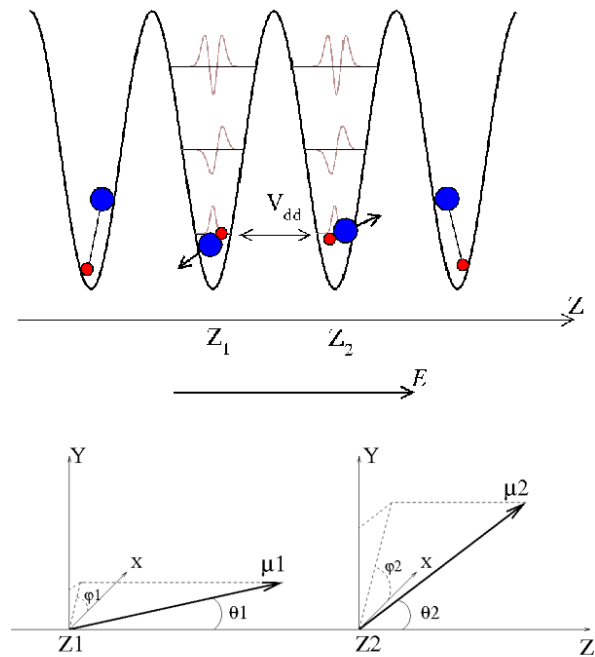}
\caption{Top figure (a): schematic of a polar molecule based quantum register \cite{DeMille2002}. Bottom figure (b): Permanent dipole moments $\mu_i$, i=1,2 of the molecules. 
Their orientations are caracterized by the angles $\theta_i$ and $\phi_i$. The electric field E is parallel to the Z-axis connecting the center of mass of the molecules. In the
simulation, the wave length of the lattice laser is $\lambda$ = 600 nm so the intermolecular distance is 300 nm, the dipole moment is 4.6 Debye and the dipole-dipole interaction is about a few tens of kilo-Hertz }
\label{register}
\end{figure}

Figure \ref{register} shows a schematic description of the polar molecule based quantum register.  It consists of a string of polar ultracold heteronuclear NaCs dimers experiencing a static electric field 
\cite{DeMille2002}. 
In the ground $X^1\Sigma^+$ electronic state, NaCs has a permanent dipole moment of 4.6 Debye at the equilibrium distance of 7.20 a.u. and is among the strongest of all alkaline mixtures \cite{Aymar2005} making it a good candidate for quantum computing. 
Photoassociation from an ultracold Na and Cs mixture can be used to obtain NaCs dimers at temperatures cold enough to be optically trapped \cite{Haimberger2004}. Many other ultracold heteronuclear dimers have been obtained using various techniques \cite{Carr2009}.
Experimental techniques for trapping and manipulating ultracold molecules are rapidly developping \cite{Carr2009}.
We assume that the molecules can be trapped in the lowest translational states of successing sites of
a three-dimensional optical lattice, with only one molecule per site and without tunneling from one site to another.
Both the string of molecules and the electric field are aligned along the Z-axis of a laboratory-fixed frame chosen to be the quantization axis.  The electric field 
orientates the molecules resulting in a mixing of the rotational levels. 
The electric field may vary along the Z-axis to make the molecules individually addressable by spectroscopic means.

Molecules are initially stored in a high vibrational level of the ground $X^1\Sigma^+$ potential. Information can be stored in these long-lived levels for they have no permanent dipole moment and the molecules can therefore be regarded as
isolated from one to another even in the presence of the electric field. When a logical operation needs to be carried out, the required molecules (depending on the 
number of qubits for the computation) are brought to the lowest vibrational levels \cite{Kuznetsova2008}. As opposed to before, these levels have a large dipole moment and the strong dipole-dipole 
interaction coupling adjacent molecules allow logic gate operations implemented on several molecules.  One should emphasize that this procedure differs from logic gate based on switchable dipole-dipole 
interactions proposed by Kuznetsova et al \cite{Kuznetsova2008} because in our case, the variation of the dipole-dipole interaction is not part of the logical operation itself. 
For molecules separated by a distance $R$ of 300 nm, the dipole-dipole interaction is of the order of a few tens of kilo-Hertz. The interaction decreases rapidly as the distance
between molecules increases due to its $1/R^3$ variation. At 500 nm the interaction magnitude drops to a few hundreds of Hertz. Desired driven transitions are differentiated from unwanted transitions by the dipole-dipole interaction. 
A small ineraction will require a longer pulse duration in order to differentiate the transitions. The separation between molecules is dictated by the caracteristics of the optical lattice the molecules
will be loaded in. We choose an intermolecular distance of 300 nm in our simulations. For NaCs, a wavelength of 600 nm corresponds to a molecular transition above the $3s+7s$ asymptote and below the inner well of the $^1\Sigma^+$ potential correlating to the $3s+7p$ asymptote \cite{Korek2000} in a region were overlap integrals with the lowest vibrational levels of the ground $^1\Sigma^+$ are extremely small offering a favorable wavelength window for an optical trap with sites distant by a distance of 300 nm.  Similar features can be found for others heteronuclear alkaline molecules \cite{Kotochigova2006}.

\subsection{Hamiltonian}

This work is restricted to the study of gates involving two or three qubits encoded in no more than two
molecules. From now on, we will use capital letters to refer to coordinates in the laboratory frame and minuscule letters 
when we refer to inner coordinates of the molecules. We note $(\mathbf{e}_X,\mathbf{e}_Y, \mathbf{e}_Z)$ the basis vector of a Cartesian space-fixed frame (or laboratory Cartesian
frame). We note $\mathbf{e}_Q, Q=0,\pm 1$ the basis vector for a space-fixed spherical coordinate system. 
We note $(\mathbf{e}_x,\mathbf{e}_y, \mathbf{e}_z)$ the basis vector of a Cartesian molecular frame and 
$\mathbf{e}_p, p=0,\pm 1$ a molecular spherical frame. When the spherical indices are ambiguous, we may use the index $\emph{R}$ and $\emph{r}$ instead of the index 0 for the laboratoy and molecular frame respectively.

We consider molecules in the ground $X^1\Sigma^+$ electronic potential only. For NaCs, we use the potential calculated as explained in \cite{Aymar2005}. The total Hamiltonian for two molecules can be written as the sum of a time independent term $H_0$ and a time dependent pertubation term $\mathbf{W}(t)$:
\begin{eqnarray}
\mathbf{H}_{tot}(t)&=&\mathbf{H}_0+\mathbf{W}(t)
\end{eqnarray}

$\mathbf{W}(t)$ represents the interaction of the molecules with a laser pulse polarized along the $\mathbf{e}_Z$ direction: $\mathbf{E}_{L}(t)=E_{L}(t).\mathbf{e}_Z$ 
\begin{equation}
\mathbf{W}(t)=\sum_i-\mathbf{\mu}_i.\mathbf{E}_{L}(t)=\sum_i-\mu_{i_0}.E_{L}(t)\cos{\theta_i}
\end{equation}
where $\mu_{i_0}$ is the $z_i=r_i$ component of the dipole moment of molecule $i$ in the ground electronic state. 
$\mathbf{H}_0$, which represents the energy of the molecules coupled by the dipole-dipole interaction, is made of several contributions:
\begin{equation}
\label{time-independent-Hamiltonian}
\mathbf{H}_0=\sum_{i=1}^{2}(\mathbf{H}_{ex}^i+\mathbf{H}_{in}^i+\mathbf{H}_{S}^i)+
\mathbf{V}_{dd}
\end{equation}

 $\mathbf{H}_{ex}^i$,  $\mathbf{H}_{in}^i$, and $\mathbf{H}_{S}^i$ are the external, internal, and Stark Hamiltonians 
associated to the molecule $i=1,2$ respectively. $\mathbf{V}_{dd}$ is the dipole dipole interaction.

The external Hamiltonian $\mathbf{H}_{ex}^i$ describes the motion of the molecule $i$ in the trapping optical potential. It can be approximated locally by a three dimensional 
isotropical harmonic oscillator with frequency 
$\omega_L$ and a depth $V_0$ \cite{Micheli2007}  
\begin{equation}
\mathbf{H}_{ex}^i=\frac{\mathbf{P}_i^2}{2M}+\mathbf{V}_{opt}(\vec{R_i}) 
\end{equation}

where $\emph{M}$ is the total mass.  $\omega_L$ and $V_0$ depend on the dynamic polarizabilities of the molecule and on the frequency and intensity of the trapping laser \cite{Micheli2007,Kotochigova2006}. Typical values for $\omega_L$ are kilohertz. The depth must be of the order of tens of kilohertz in order for the molecules to be tightly trapped and for the tunneling from one site to another to be negligible. 

The inner Hamiltonian $\mathbf{H}_{in}^i$ describes an isolated diatomic molecule in the Born Oppenheimer approximation. It contains the vibrational Hamiltonian $\mathbf{H}_{vib}^i$
in the ground $X^1\Sigma^+$ electronic state and the rotational Hamiltonian $\mathbf{H}_{rot}^i$: 
\begin{equation}
\mathbf{H}_{in}^i=\mathbf{H}_{vib}^i+\mathbf{H}_{rot}^i 
\end{equation}
  with
\begin{equation}
\mathbf{H}_{vib}^i=\frac{\mathbf{p}_{i}}{2m}+\mathbf{v}(r_i) 
\end{equation}
 where $m$ is the reduced mass of the molecule vibrating in the interatomic potential $\mathbf{v}(r_i)$, $r_i$ being the interatomic coordinate.

For $^1\Sigma$ molecules, the rotational Hamiltonian for a rigid rotator is given by:
\begin{equation} 
\textbf{H}_{rot}^i=B_v^i\textbf{N}^2
\end{equation}
where $\textbf{N}$ is the total angular momentum and $B_v^i$ is the rotational constant in vibrational state $v$.

$\mathbf{H}_{S}^i$ refers to the Stark Hamiltonian of a molecule in a static electric field $\mathbf{E}_S^i(\vec{R_i})$. In our case, $\mathbf{E}_S^i(\vec{R_i})$ is aligned along the $Z$ axis, and varies only along this coordinate. 
 The Stark Hamiltonian reduces to:
\begin{equation} 
\label{stark-final}
\mathbf{H}_{S}^i=-E^i(Z_i)\mu_{i_0} \cos(\theta_i).
\end{equation}

The last contribution comes from the long range dipole-dipole interaction term $\mathbf{V}_{dd}(\mathbf{R})$ of two polar molecules separated 
by a distance $\mathbf{R}=\mathbf{R}_2-\mathbf{R}_1=Re_Z=Re_R$   in the laboratory frame. By assuming that the molecules rotate in a plan, the projection of the rotational quantum number on the internuclear axis is conserved and the dipole-dipole interaction term becomes \cite{Charron2007}: 
\begin{equation}
 \mathbf{V}_{dd}(R,\theta_1\theta_2)=-\frac{1}{2\pi\epsilon_0}\frac{\mu_{0,i}\mu_{0,j}}{R^3}\cos{\theta_1} \cos{\theta_2}
\end{equation}

where the angles $\theta_1$ and $\theta_2$ caracterize the orientation of the dipole moments versus the intermolecular axis $\mathbf{e}_R$.

\subsection{Product and coupled basis sets}
For a system of 
two interacting molecules, we define a product basis set $|\varphi\rangle$ constructed by tensorial product of the individual basis of the individual molecules. 
\begin{eqnarray}
|\varphi \rangle&=&|n,v,N,m_N \rangle_1\otimes|n,v,N,m_N\rangle_2 \\
&=&|n_1,n_2,v_1,N_1,m_{N_1},v_2,N_2,m_{N_2} \rangle
\end{eqnarray}
where the individual basis set $|n,v,N,m_N\rangle_i$ is built from a tensor product of the eigenbasis of $\mathbf{H}_{ex}^i$, $\mathbf{H}^i_{vib}$, and $\mathbf{H}^i_{rot}$. 

We make the assumption 
that $\mathbf{V}_{dd}$ and $\mathbf{H}_S$ only couple rotational levels. However, both terms can create couplings between states of the optical trap. For a molecule 
in the ground state of the trap, the translational ground state wave-function of the potential well is a Gaussian of width $a=\sqrt{\hbar/M\omega_L}$. For a typical experiment, 
$a$ is of the order of tens of micro-meters. No coupling will occur if $\mathbf{H}_S$ is constant over the range of variation of $a$, the coupling integral $\langle n_i| \mathbf{H}_S |m_i \rangle$  involving the eigenfunctions 
of the harmonic trap being zero in this case. For the couplings due to the dipole-dipole interaction, a detailed analysis can be found in ref \cite{Kuznetsova2008}. Coupling between translational states for typical 
polar molecules is small and results in an energy shift no larger than 1\% of the unperturbed energy spacing. For typical optical trap, where the spacing between 
translational levels is about tens of kilo-Hertz, it gives a negligeable shift of a few tens of Herz. the $n$ label can then be omitted in the basis function notation.

Also, we only consider motion in a plane $\phi_i = 0$ with a field aligned along $\mathbf{e}_Z$ so only $m_{N}=0$ states are involved. Finally, the relevant contracted notation
 $|v_1,N_1,v_2,N_2\rangle$ will label the states of the total product basis. It can be further simplified to just $|N_1,N_2\rangle$ when different vibrational states are not considered. 

The states of the coupled basis set that diagonalizes the full time independent Hamiltonian $\mathbf{H}_0$ are
noted $|\tilde{\varphi}\rangle=|\tilde{v}_1,\tilde{N}_1,\tilde{v}_2,\tilde{N}_2\rangle$ (or $|\tilde{\varphi}\rangle=|\tilde{N}_1,\tilde{N}_2\rangle$ when the vibrational manifold is well known) to indicate 
the product basis state $|v_1,N_1,v_2,N_2\rangle$.to which they are adiabatically connected.

\subsection{Dynamics}

The time dependent Schr\"odinger equation will be solved in the coupled basis set in the interaction representation \cite{Cohen-Tannoudji}.
The evolution of the populations $b_n(t)$ are given by the set of coupled equations :
\begin{equation}
 i\hbar\frac{d}{dt}b_n(t)=\sum_k e^{i\omega_{nk}t}W_{nk}(t)b_k(t)
\end{equation}
where we have introduced the Bohr pulsation $\omega_{nk}=(E_n-E_k)/\hbar$ and the matrix elements $W_{nk}(t)$ of the time-dependent interaction in the eigenbasis of $H_0$. 
\begin{equation}
W_{nk}(t)=\langle\tilde{\varphi}_n|W|\tilde{\varphi}_k \rangle
\end{equation}
The coupled equations are solved using the fourth-order Runge-Kutta method \cite{NumericalRecipes} without any rotating wave approximation.

\begin{figure}[h]
\includegraphics[width=0.9\columnwidth]{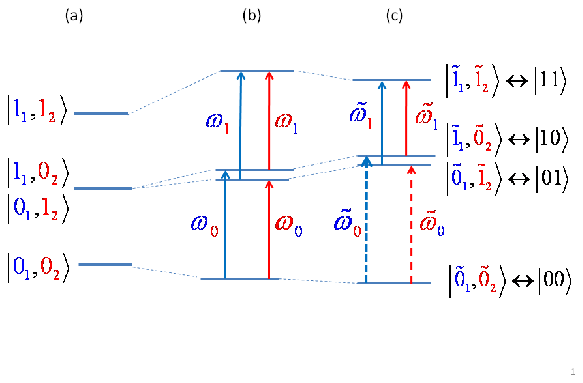}
\caption{Diagram of the lowest eigenvalues of the total time-independent Hamiltonian for three different cases: without any electric field (a), with electric fields $E^1>E^2$
without and with dipole-dipole interaction ((b) and (c) respectively). The eigenvectors $|\tilde{N}_1,\tilde{N}_2 \rangle$ correlate adiabatically to the vector $|N_1,N_2 \rangle$ with
$v_1=v_2=0$. Only $m_N=0$ are considered. the $m_N$ anf $v_i$ labels are omitted. Transitions in the first or second molecule are shown by blue or red arrows respectively. 
The subscripts labelling the frequencies indicate the rotational state of the neighboring molecule.}
\label{Stark_Scheme}
\end{figure}

\section{Intermolecular controlled-NOT gate}
\label{conditional}

We first examine the possibility of steering elementary gates by $\pi$ pulses in the microwave and infrared domain by focusing on conditional population inversions. The control of the phase will be considered in the final section. 

The first example concerns the two-qubit controlled-NOT (CNOT) gate which flips the second (target) qubit if the first one (control qubit) is equal to 1. 

\begin{equation}
\begin{split}
 (\alpha|00\rangle+\beta|01\rangle+\gamma|10\rangle+\delta|11\rangle)\xrightarrow{CNOT}\\
(\alpha|00\rangle+\beta|01\rangle+\gamma|11\rangle+\delta|10\rangle).
\end{split}
\end{equation}
The gate is driven by a $\pi$-pulse defined by
\begin{equation}
 E_{\pi}(t)=E^0_{\pi}sin^2\left( \frac{\pi t}{\tau_p} \right) cos(\omega_{if} t)
\end{equation}
with amplitude $E^0_{\pi}=\frac{2\pi}{\tau_p \mu_{if}}$ where $\index{\footnote{}}\tau_p$ is the duration of the pulse and $\mu_{if}$ is the dipole moment for 
the transition. The scheme is similar to some early proposals for conditional quantum dynamics \cite{Barenco1995,Berman1997}.
It uses the fact that a transition frequency in the second molecule depends on the state of the first molecule. 

The qubits are encoded in two neighboring molecules.  Both molecules are in the ground vibrational states 
$\tilde{v}_1=\tilde{v}_2=0$ so that the states are denoted only by
the rotational states $|\tilde{N}_i,\tilde{N}_j\rangle$. The logical states are encoded into the first four states of
the coupled basis set which correlates with the $N_1=0,1$ and $N_2=0,1$ states (see Fig. \ref{Stark_Scheme}).
In the product basis the CNOT transformation should consist in flipping the rotational state of the second molecule if and only if the first one is in state $N_1=1$.
In a basis set diagonalizing the Stark Hamiltonian without dipole-dipole interaction, the frequencies of the rotational transitions in the second molecule when the first one is in the state $N_1=0$ ($\omega_{0}$) or
$N_1=1$ ($\omega_{1}$) are the same (see Fig.\ref{Stark_Scheme}). The dipole-dipole interaction leads to two very close but different frequencies. This allows us to excite only 
the transition $|\tilde{1}\tilde{0}\rangle \rightarrow|\tilde{1}\tilde{1}\rangle$ at a frequency $\tilde{\omega}_1$ (full red arrow in Fig. \ref{Stark_Scheme}) and not the transition
$|\tilde{0}\tilde{0}\rangle\rightarrow|\tilde{0}\tilde{1}\rangle$  at a frequency $\tilde{\omega}_0$ (dashed red arrow). When the system is in a computational basis state, a $\pi$-pulse can
selectively drive the desired population inversion. To get a perfect selectivity and avoid the off resonance transition the pulse duration must satisfy the condition
$\tau_p>10/|\tilde{\omega}_1-\tilde{\omega}_0|$
\cite{Vasilev2004}. 

\begin{figure}
\includegraphics[width=0.9\columnwidth]{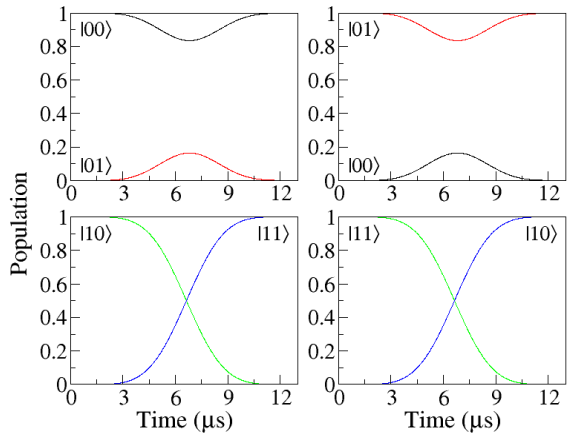}
\caption{Evolution of the populations during the CNOT gate driven by a $\pi$-pulse. When the control qubit is 0 (initial states $|00\rangle$ or 
$|01\rangle$), the final state of the qubits remains unchanged when the pulse is applied although the population may vary during the process (top left and right panels respectively). 
On the contrary, the second qubit is flipped when the initial state is  $|10\rangle$ or $|11\rangle$ leading to final states $|11\rangle$ or $|10\rangle$(bottom left and right panels respectively).}
\label{CNOTPI}
\end{figure}

Fig.\ref{CNOTPI} shows the evolution of the populations during the CNOT gate driven by a $\pi$-pulse starting from the four different states of the computational basis. The states corresponding to the logical 
states $|00\rangle$ and $|01\rangle$ temporarily leave the
computational space but correctly return to it with an arbirary phase (the phase constraint will be discussed after). From now on, simulations are ran with the following fixed parameters: $R_{12}=$ 300 nm, $E^1=$ 2kV.cm$^{-1}$, $E^2 = $1.5 kV.cm$^{-1}$. 
The dynamical basis set for the simulation contains the vibrational state $v=0$ and rotational states up to $N=4$ for each molecule. The difference of frequencies which must be
resolved is 4.002$\times 10^{-6}$ cm$^{-1}$ and the pulse duration must be $\tau_p=13.2$ $\mu$s. 
The intensity of the $\pi$-pulse is $E_{\pi}^0= 7.5$ V.m$^{-1}$. 
 . 

\section{0 and 1 ADDER}

\begin{figure}
\includegraphics[width=0.9\columnwidth]{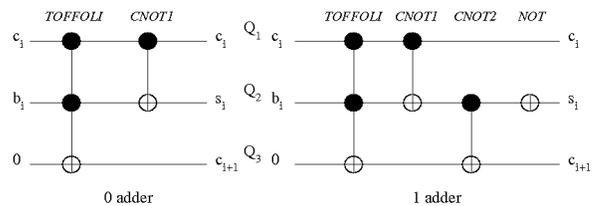}
\caption{Logical circuits for the 0 adder (left panel) and 1 adder (right panel).}
\label{logical-circuits}
\end{figure}

The adder of 0 and 1 (0-ADD and 1-ADD) \cite{Beckman1996} is a simplification of the more general full adder of two binary digits $a_i$ and $b_i$ and a carry in $c_i$
\cite{Vedral1996}. Similarly to the classical case, the addition of two numbers in binary representation, $a=(a_1a_2...a_i...a_n)$ and $b=(b_1b_2...b_i...b_n)$ is performed digit by digit 
starting from the least significant one. For the $i^{th}$ rank, one needs to evaluate a sum
$s_i$ and a carry out $c_{i+1}$. A cycle of the quantum full adder requires four qubits \cite{Vedral1996, Benenti}. 
However, in the special case of a 0 or 1 adders, only three qubits operations are necessary. The $c_i$ and $b_i$ digits are encoded in the first two qubits and the third is 0. One
 has initially 
$|Q_1Q_2Q_3\rangle = |c_i,b_i,0\rangle$. The information that $a_i$ is 0 or 1 repectively is contained in the applied pulse which is different for the 0-ADD or 1-ADD gate. 
The results  $s_i$ and $c_{i+1}$ are stored in
the second and third qubits respectively. One has finally $|Q_1Q_2Q_3\rangle = |c_i,s_i,c_{i+1}\rangle$. The 0-ADD or 1-ADD drives the unitary transformation 
$|c_i,b_i,0\rangle \rightarrow  |c_i,s_i,c_{i+1}\rangle$ for $a_i$ is 0 or 1 repectively. The truth tables are given in Table \ref{truth-table} (note that only the first four 
inputs with $Q_3=0$ are effectively used, the last four ones insure the reversibility of the operations). 

As schematized in Fig.\ref{logical-circuits}, the carry out of
the 0-ADD gate is  $c_i$.AND.$b_i$ and is computed by the
three-qubit TOFFOLI gate: $|x\rangle|y\rangle|z\rangle\rightarrow |x\rangle|y\rangle|z\oplus x.y\rangle$ where $\oplus$ is the sum modulo 2. 
The TOFFOLI gate corresponds to the AND gate when the
third qubit is zero. It is the controlled-controlled NOT which flips the third qubit if the first two qubits are in state 1.  
The sum $s_i$ is given by the first CNOT gate (noted CNOT1): $|x\rangle|y\rangle\rightarrow |x\rangle|y\oplus x\rangle$  The 1-ADD is the continuation of the 0-ADD. 
For the 1-ADD, we can see from Table \ref{truth-table} that $s_i$ is
obtained from the $s_i$ of the 0-ADD by an additional NOT gate which simply flips the state ($|0\rangle$ and $|1\rangle$) of the qubit. It can be seen also that $c_{i+1}$ of the 1-ADD only needs to do the sum modulo 2 of the $s_i$ and the carry out
$c_{i+1}$ of the 0-ADD which are encoded in the second and third qubits. As a result this step is a new CNOT (noted CNOT2) with control qubit $Q_2$ and target qubit $Q_3$.

The most promising implementation of the 0-ADD and 1-ADD consists in encoding the carry $c_i$ and the number $b_i$ in the rotational and the vibrational structure of the first molecule whereas
the carry out $c_{i+1}$ is encoded in the rotation of the second molecule. So one has $|Q_1Q_2Q_3\rangle =|\tilde{N}_1,\tilde{v}_1,\tilde{N}_2\rangle$.
The sum $s_i$ for the $i^{th}$ step of the addition will be stored in the vibration of the first molecule in replacement of $b_i$ and the carry out encoded in the second
molecule becomes the
carry in for the $(i+1)^{th}$ step allowing the addition to be further continued. 

When the logical states are mapped
on a coupled basis set and not on a product basis set, one has to steer the transformation in the total computational basis set (a three-qubit space here). For a one-qubit (or two-qubit) transformation one has to consider all the transformations involving the active qubit(s)
 in the 3-qubit space. For example, the extended operator of the NOT gate on $Q_2$ is  
\begin{equation}
\tilde{U}_{Q_2}=E_{Q_1}\otimes U_{Q_2}\otimes E_{Q_3}
\end{equation}
where $E_{Q_k}$ is the unity matrix in the one-qubit space $k$. This leads to four population inversions in place of a single one in a product basis set. Secondly, one should point out that whatever the encoding is, unwanted
transitions can always occur. The length of the pulses are then chosen to satisfy two opposite
conditions. The pulse must be short enough to simultaneously drive a maximum number of active transitions with a single carrier frequency and it must be long enough to avoid
these unwanted transitions whose frequencies are very close.

\begin{table}
\begin{tabular}{ccc|ccc||ccc|ccc}
\hline
\multicolumn{6}{c||}{0 adder}  & \multicolumn{6}{c}{1 adder} \\
c$_i$&b$_i$&Q$_3$&c$_i$&s$_i$&c$_{i+1}$ &c$_i$&b$_i$&Q$_3$&c$_i$&s$_i$&c$_{i+1}$ \\
0&0&0&0&0&0 & 0&0&0&0&1&0 \\
0&1&0&0&1&0 & 0&1&0&0&0&1 \\
1&0&0&1&1&0 & 1&0&0&1&0&1 \\
1&1&0&1&0&1 & 1&1&0&1&1&1 \\
&&&&&&&&&&& \\
c$_i$&b$_i$&Q$_3$&c$_i$&s$_i$&\={c}$_{i+1}$ &c$_i$&b$_i$&Q$_3$&c$_i$&s$_i$&\={c}$_{i+1}$ \\
0&0&1&0&0&1 & 0&0&1&0&1&1 \\
0&1&1&0&1&1 & 0&1&1&0&0&0 \\
1&0&1&1&1&1 & 1&0&1&1&0&0 \\
1&1&1&1&0&0 & 1&1&1&1&1&0 \\

\end{tabular}
\caption{0-adder and 1-adder truth tables. The information $a_i =0$ and $a_i =1$ respectively is contained in the pulse driving the transformation. Only the first four 
inputs with $Q_3=0$ are effectively used. \={c}$_{i+1}=1\oplus c_{i+1}$. }  
\label{truth-table}
\end{table}

\subsection{Intermolecular TOFFOLI gate}

\begin{figure}
\includegraphics[width=0.9\columnwidth]{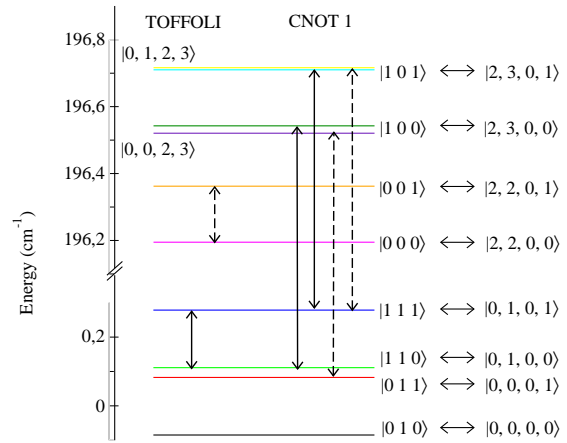}
\caption{Encoding of the three qubits for the 0 ADDER in the states of the coupled basis set $|Q_1Q_2Q_3\rangle \leftrightarrow |\tilde{v}_1,\tilde{N}_1,\tilde{v}_2,\tilde{N}_2\rangle$. The
active transitions for the TOFFOLI and CNOT gates (the latter is called CNOT1 in the text) are shown in full arrows and the corresponding unwanted transitions are shown by dashed
arrows. The zero of energy is chosen at the ground state without Stark field and dipole-dipole coupling. }
\label{adder_encoding}
\end{figure}

Conditional dynamics are more demanding for the TOFFOLI gate than for
the CNOT gate presented in the previous section. Although
the computional basis $|Q_1Q_2Q_3\rangle$ generated by the eight combinations of $\tilde{N}_1=0,1$, $\tilde{v}_1=0,1$, $\tilde{N}_2=0,1$ (with $\tilde{v}_2$ always equal to 0)
would provide an intuitive mapping between the quantum numbers and the logical states, it is rather unconveniant in practice because of the very similar
rotational constants for the two vibrational levels. With this assigment, the rotational state of the second molecule must be flipped if and only if the first molecule is 
in a given rotational and vibrational state.  Then, except for $\tilde{N}_2$ holding the
carry out, we adopt an assignment with no correspondance between the quantum numbers and the associated qubit state. The eight states and the logical mapping are shown in Fig.\ref{adder_encoding}.  In the present case, the frequency of the
$|110\rangle\leftrightarrow|111\rangle$ TOFFOLI transition (solid arrow in Fig.\ref{adder_encoding}) is fairly close to the unwanted $|000\rangle\leftrightarrow|001\rangle$
transition. The difference in frequency which must be resolved is then $4.0\times10^{-7}$ cm$^{-1}$. 

The population of each computational basis state 
during the TOFFOLI gate is plotted in the top panel (a) of Fig. \ref{0adder_pop.eps}. Each population is weighted by a factor different from 1 for clarity (it is not a
superposed state). The duration of the $\pi$ pulse is $\tau_p$ = 253 $\mu$s. The carrier frequency is in resonance with the active transition 
$|110\rangle\leftrightarrow|111\rangle$, i.e. $|\tilde{0},\tilde{1},\tilde{0},\tilde{0}\rangle\leftrightarrow|\tilde{0},\tilde{1},\tilde{0},\tilde{1}\rangle$ with 
$\omega$= 0.16743 cm$^{-1}$. The amplitude of the pulse is $E_0=4.1506\times10^{-6}$ kV/cm.
For all the simulations relative to the adder algorithm, the dynamical basis set contains 64
states including the $v_2 = 2$ manifold ($v_1=v_2=0$ with $J_1$ and $J_2 = 0,1,2$; $v_1=v_2=2$ with $J_1=0,1,2,3,4$).

The population inversion is very good with a fidelity of 0.9999. Note that the $\pi$ pulse is not optimized for a superposition but is sufficient for arithmetic operations for
which the system is always in a computational basis state. As illustrated in the next section, optimization for
any superposition can be obtained with optimal control theory.

\subsection{Intramolecular CNOT1 gate}

 This gate is intramolecular since both the control $Q_1$ and target $Q_2$
qubits are encoded in the first molecule (the rotation and the vibration respectively).The extended CNOT1 gate involves the following 
transitions: $|10Q_3\rangle\leftrightarrow |11Q_3\rangle$ for any value of the third qubit $Q_3$. These two transitions are shown in full arrow in Fig. \ref{adder_encoding}: 
\begin{eqnarray}
\omega_{Q_3=0}=|100\rangle\equiv|\tilde{2}\tilde{3}\tilde{0}\tilde{0}\rangle&\leftrightarrow&|110\rangle\equiv|\tilde{0}\tilde{1}\tilde{0}\tilde{0}\rangle \\
\omega_{Q_3=1}=|101\rangle\equiv|\tilde{2}\tilde{3}\tilde{0}\tilde{1}\rangle&\leftrightarrow&|111\rangle\equiv|\tilde{0}\tilde{1}\tilde{0}\tilde{1}\rangle
\end{eqnarray}
They correspond to vibrational transitions in the first molecule and thus belong to the infrared domain.
The two transitions being very close ($\Delta\omega = 8.3\times10^{-7}$ cm$^{-1}$), they can be induced by a 
single $\pi$-pulse. Two unwanted frequencies are shown by dashed arrows
in Fig.(\ref{adder_encoding}). They involve vibrational states of the second molecule. Fortunately, these transitions differ from the active transitions by about 
$6\times10^{-3}$ cm$^{-1}$. A shorter pulse than for the TOFFOLI gate can be used. 

\begin{figure}
\includegraphics[width=0.9\columnwidth]{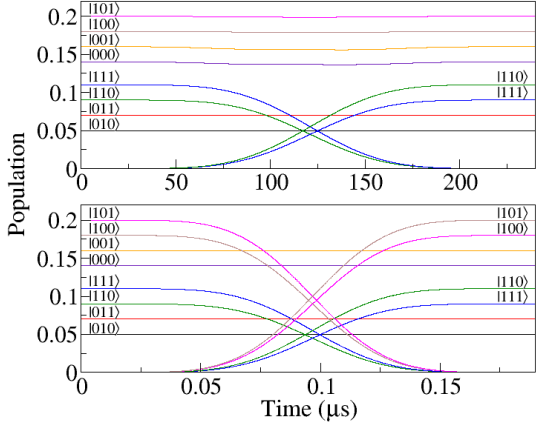}
\caption{Population evolution for each computational basis state during the gates of the 0 ADDER. Upper panel: TOFFOLI gate. Lower panel: CNOT1 gate. Each population is
weighted by a factor different from 1 for clarity (it is not a superposed state)}
\label{0adder_pop.eps}
\end{figure}

Botton panel (b) of Fig.\ref{0adder_pop.eps} shows the populations of each computational basis state during the CNOT1 gate.
The pulse duration is 
$\tau_p=0.194$ $\mu$s.  The carrier frequency is fixed by the $|100\rangle\leftrightarrow |110\rangle$ transition and is equal to
$\omega_{Q_3=0}= 196.43$ cm$^{-1}$. The amplitude is $E_0=93.793$ kV/cm.

\subsection{Intermolecular CNOT2 gate}

 This logical operation is $|1Q_{2}0\rangle\leftrightarrow |1Q_{2}1\rangle$ for any value of
the second qubit $Q_2$. We then have two active transitions 
\begin{eqnarray}
\omega_{Q_2=0}=|100\rangle\equiv|\tilde{2}\tilde{3}\tilde{0}\tilde{0}\rangle\leftrightarrow|101\rangle\equiv|\tilde{2}\tilde{3}\tilde{0}\tilde{1}\rangle \\ 
\omega_{Q_2=1}=|110\rangle\equiv|\tilde{0}\tilde{1}\tilde{0}\tilde{0}\rangle\leftrightarrow|111\rangle\equiv|\tilde{0}\tilde{1}\tilde{0}\tilde{1}\rangle 
\end{eqnarray}

Top panel (a) of Fig.\ref{1adder_pop.eps} shows the evolution of populations corresponding to the intermolecular CNOT gate (named CNOT2).
A superposition of two microwave $\pi$-pulses are used. 
As for the TOFFOLI gate, we need to resolve a frequency difference of $4\times10^{-7}$cm$^{-1}$
The parameters for the pulses are: $\tau_p = 242 \mu s$
$\omega_{Q_2=0}=0.16734$ cm$^{-1}$, $E_0=4.1509\times10^{-6}$ kV/cm and $\omega_{Q_{2=1}}= 0.16744$ cm$^{-1}$ $E_0=4.1508\times10^{-6}$ kV/cm.

\subsection{Intramolecular NOT gate}

 The frequencies of the four active transitions
of the extended NOT gate ($|Q_{1}0Q_{3}\rangle\leftrightarrow |Q_{1}1Q_{3}\rangle$ for any values of $Q_1$ and $Q_3$) are 
 \begin{eqnarray}
\omega_{Q_1=0,Q_3=0}&=|000\rangle\equiv|\tilde{2}\tilde{2}\tilde{0}\tilde{0}\rangle\leftrightarrow|000\rangle\equiv|\tilde{0}\tilde{0}\tilde{0}\tilde{0}\rangle \\
\omega_{Q_1=1,Q_3=0}&=|100\rangle\equiv|\tilde{2}\tilde{3}\tilde{0}\tilde{0}\rangle\leftrightarrow|110\rangle\equiv|\tilde{0}\tilde{1}\tilde{0}\tilde{0}\rangle \\
\omega_{Q_1=0,Q_3=1}&=|001\rangle\equiv|\tilde{2}\tilde{2}\tilde{0}\tilde{1}\rangle\leftrightarrow|011\rangle\equiv|\tilde{0}\tilde{0}\tilde{0}\tilde{1}\rangle \\
\omega_{Q_1=1,Q_3=1}&=|101\rangle\equiv|\tilde{2}\tilde{3}\tilde{0}\tilde{1}\rangle\leftrightarrow|111\rangle\equiv|\tilde{0}\tilde{1}\tilde{0}\tilde{1}\rangle
\end{eqnarray}

\begin{figure}
\includegraphics[width=0.9\columnwidth]{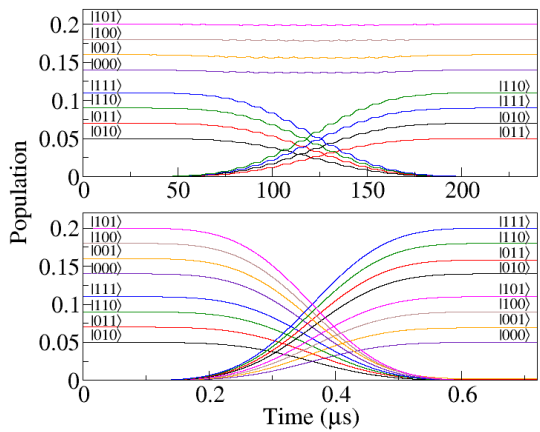}
\caption{Population evolution of each computational basis state  during the supplementary gates of the 1 ADDER. Upper panel: CNOT2 gate. Lower panel: NOT gate.Each population is
weighted by a factor different from 1 for clarity (it is not a superposed state)}
\label{1adder_pop.eps}
\end{figure}

Bottom panel (b) of Fig.\ref{1adder_pop.eps} shows the population evolutions for the final NOT gate. 

The frequencies $\omega_{Q_1=0,Q_3=0}$ and $\omega_{Q_1=1,Q_3=0}$ on one side and $\omega_{Q_1=0,Q_3=1}$ and $\omega_{Q_1=1,Q_3=1}$ on the other side are sufficiently close to be driven by a single
$\pi$-pulse ($\Delta\omega=4\times10^{-6}$ cm$^{-1}$ and $\Delta\omega=8\times10^{-7}$ cm$^{-1}$ respectively). The two pulses have a duration adapted to avoid unwanted
transitions ($\Delta \omega = 6\times10^{-3}$ cm$^{-1}$), $\tau_p = 0.726 \mu s$. The frequencies are $\omega_{Q_1=0,Q_3=0}=196.28$ cm$^{-1}$ and
$\omega_{Q_1=1,Q_3=0}=196.43$ cm$^{-1}$. The amplitudes are $E_0=25.0115$ kv/cm and $E_0=14.4558$ kv/cm.

\subsection{Intramolecular initialization}

 Finally, one has to discuss the initialization of an addition cycle. The digit $b_i$ must be encoded in the first molecule for any value of the carry in $c_i$ which is unknown.
With the assignment used here (see Fig.\ref{adder_encoding}), the digit is encoded in the vibration of the first molecule. If $b_i=1$, the molecule must be in the manifold $\tilde{v}_1=0$. This means that the system is directly ready since the algorithm generates the carry out in this ground vibrational state. If $b_i=0$, the molecule must be in the manifold $\tilde{v}_1=2$. Two transitions must then be driven:
\begin{eqnarray}
|010\rangle\equiv|\tilde{0}\tilde{0}\tilde{0}\tilde{0}\rangle &\rightarrow&|000\rangle\equiv|\tilde{2}\tilde{2}\tilde{0}\tilde{0}\rangle\\
|110\rangle\equiv|\tilde{0}\tilde{1}\tilde{0}\tilde{0}\rangle&\rightarrow&|100\rangle\equiv|\tilde{2}\tilde{3}\tilde{0}\tilde{0}\rangle
\end{eqnarray}
They correspond to two transitions with frequencies $\omega_{Q_1=0,Q_3=0}$ and $\omega_{Q_1=1,Q_3=0}$  of the NOT gate realized above and the NOT pulse can then be used to initialize the addition cycle.

\subsection{Concatenation}

\begin{figure}
\includegraphics[width=0.9\columnwidth]{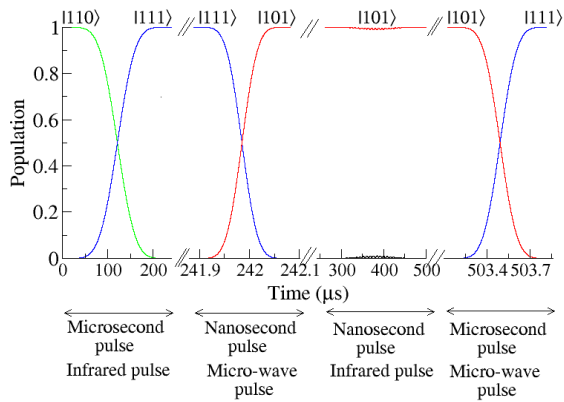}
\caption{Population evolution of the computational basis state corresponding to the logical state $|110 \rangle$ ($b_i=1$ and $c_i=1$ during the  1-ADDER gate
 (TOFFOLI-CNOT1-CNOT2-NOT gates) to give the final logical state $|111 \rangle$ corresponding to $s_i=1$ and $c_{i+1}=1$. }
\label{conca}
\end{figure}

 Fig. \ref{conca} illustrates the full 1-ADDER operation for the example $b_i=1$ and $c_i=1$. The initial logical state is $|110 \rangle$. The network TOFFOLI (intermolecular), CNOT1 (intramolecular), CNOT2 (intermolecular) and NOT (intramolecular) gates operating in the microwave or infra red domain drives the system towards the
final  $|111 \rangle$ logical state corresponding to $s_i=1$ and $c_{i+1}=1$.

\subsection{Discussion}

Any perturbation to the energy level structure coming from neighboring molecules, hyperfine interactions, coupling with the translational states in the lattice will drastically 
affect the high fidelity logical operations the pulses were designed to drive. 

 We chose the electric field to be very different from one molecule to the other (2kV/cm and 1.5kV/cm in our simulations). 
The reason for this choice was motivated by the fact that we have intramolecular gates implemented on one single molecule and well differentiated molecules allows for fast 
intramolecular gates. 
Having such a large variation of the electric field may generate couplings between translational states of the trapping potential because 
the requirement of a constant field over the range of variation of $a$ might be experimentally challenging to achieve. A much smoother gradient might prove to be more 
suitable to this respect. 
We tried simulations with different gradients and it was always possible to drive algorithm although intramolecular gate times had to be longer when the gradient was small ($E^1=2$ kV/cm and $E^2=$ 1.9kV/cm). On the contrary, the time scale for intermolecular gate remains the same being mainly fixed by the dipole-dipole interaction. 

We assumed that the electric field was orientated along the intermolecular axis. Note that results will remain valid for other field orientations. Only the expression of the interaction will differ from what we presented in this model, but the order of magnitude of the interaction will be about the same.

The effect of neighboring molecules interacting on the two-molecule systems has been evaluated. 
We have analyzed the energy shifts on the levels of the active molecules $j$ and
$j + 1$ by adding two molecules (at position $j - 1$ and $j + 2$) and using the same gradient for the Stark field as for our simulations. 
The basis set is composed of manifolds $v = 0$ with $N = 0,1,2$ and $v = 2$ with $N
= 0,1,2,3,4$ for each molecule. The shifts indroduced are of the order of a few kilo-Hertz. the effect is only a shift in frequency but the general structure of the schemes is not affected. Inter and intramolecular gates could still be driven with similar pulses in a larger network.

 The hyperfine structure of rovibrational molecules can complicate the manipulation of molecules with microwave pulses 
because unwanted transitions can occur \cite{Aldegunde2009,Ran2010}.
In the present model, the hyperfine structure was not taken into account. 
Within the rotational level $N=0$ manifold, the most important term is the scalar hyperfine coupling proportional to 
$\mathbf{i}_1.\mathbf{i}_2$. where $\mathbf{i}_1$ and $\mathbf{i}_2$ are the nuclear spins of the atoms of the molecule. 
The splitting between hyperfine levels is typically of the order of a kilo-Hertz. For $N$ different than 0, the most important contribution to the hyperfine Hamiltonian comes 
from the nuclear quadrupolar interaction which gives a splitting of the order of a few hundreds of kilo-Herts. In our case, $i_{Na}=3/2$, and $i_{Cs}=7/2$. $N$=0 splits into 32 
states, and $N$=1 splits into 96 levels. For a static electric field of 2kV/cm, more that 4 rotational levels per vibrational manifold are needed to described accurately the energy structure. 
Taking into account all the hyperfine levels would complicate the theoretical simulations since we had to manipulate the tensor product of the individual basis. 
However, the hyperfine structure won't change the conclusion of this work. 
 Simulations of the logical operations showed two different caracteristic time scales for the logic gates. 
Fast gates can be driven with pulses no longer than a few tens of ns. These pulses are broad enough that all hyperfine transitions within one rotational level will be excited equally, and the unresolved hyperfine structure can be neglected. 
On the other hand, slower gates such as the TOFFOLI of the adder algorithm are driven by much narrower pulses, and the splitting between hyperfine levels, larger than the dipole-dipole interaction splitting, will be well resolved and only one hyperfine level will be active. 
We note as well that as the static electric field increases orientating the molecules along the electric field, $m_N$ becomes a fairly good quantum number, 
and the matrix of the transition dipole moments between hyperfine levels of different rotational manifold show that for electric field typically stronger than 1kV/cm, only a few transitions are allowed reducing considerably the possibilities for unwanted transitions. 

Other terms that we have not considered are the second order Stark effects resulting from the static electric and the optical trapping potential. The static second order Stark 
effect is of the order of a few kilo-Hertz for the diagonal terms. The effect due to the laser field can be more important due to the wavelength dependence of the molecular
polarizabilities. For a laser intensity typically of the order of 1kW/cm$^2$, and a molecular polarizabilty of a few thousands of atomic units, the second order Stark effect is of
the order of one hundred of kilo-Hertz for the diagonal terms. It is not strong enough to affect our results.

The numerical simulations suggest that ultra cold trapped polar molecules are promising for the concatenation of  several gates with a high fidelity. First, we can 
compare the efficiency of a couple of diatomic systems with a tetraatomic molecules to implement arithmetic operations. It was not easy to encode the four qubits of the full adder
 in the two interacting dimers. Some hundreds of microseconds to add 1 is finally very longer than the timescale of the full 
addition in a polyatomic entity (some tens of ps). 
 However the main point is the possibility to avoid intermediary reading out and re-encoding of the carry out. Here the 
system is directly ready for the next cycle involving the next 
 molecule. This is a step towards scalability but at the price of long pulse duration. The other advantage is the simplicity 
and robustness of the scheme based on $\pi$ pulses. \\

\section{Phase correct gates}
\label{sec:phase}

The relative phase of the gate transitions optimized among the states of the computational basis set can reduce the accuracy of the gate when it is applied to an arbitrary
superposed state.  $\pi$ pulses are not sufficient to give phase-correct quantum gates. Then correct gate pulses can be determined by
the optimal control theory generalized to the multi-target case \cite{Tesch2002} with a phase constraint \cite{Tesch2004,Zhao2006}.

\subsection{Multi-target optimal control}

The optimal field maximizes the objective functional $J$ with the constraint that the Schr\"odinger equation
is satisfied at any time \cite{Zhu1998b,Ohtsuki2007}. The functional reads: 
\begin{widetext}
\begin{equation}
J=\sum_{n=1}^Z |\langle \psi_i^n(t_f)|\phi_f^n \rangle |^2 
- 2 \Re \left[  \int_0^{t}\langle \psi_f^n(t)| \partial_t+\frac{i}{\hbar}\mathbf{H} |\psi_i^n(t) \rangle \right]  
- \alpha \int_0^{t_f}E^2(t)dt
\end{equation}
\end{widetext}
where $\alpha$ is a positive penalty factor chosen to weight the importance of the laser fluence. For a N-qubit
gate, $Z=2^N+1$ where $2^N$ is the number of input-output transitions in the gate transformations
and the supplementary equation is the phase constraint. The $\psi_i^n(t)$ are the wave packets which are
propagated forwards in time with the initial conditions $\psi_i^n(t=0)=\phi_i^n$, $n=1,...Z$. The Lagrange
multipliers $\psi_f^n(t)$ are propagated backwards in time with the final conditions $\psi_f^n(t=t_f)=\phi_f^n$, $n=1,...Z$.  The supplementary transfer which imposes the phase correction is a sum
over all the transitions of the gate \cite{Tesch2004}
\begin{equation}
 \frac{1}{\sqrt{2^N}} \sum_{k=1}^{2^N} |\psi_i^k\rangle \rightarrow \frac{1}{\sqrt{2^N}}\left[\sum_{j=1}^{2^N} |\psi_f^k\rangle \right]e^{i\phi}
\end{equation}
and the single phase $\phi$ can take any value between 0 and $2\pi$. The universal gate field is a sum of Z contributions
\begin{equation}
E_j(t)=-(s(t)/\hbar\alpha)\Im \left[ \sum_{n=1}^Z \langle\psi_f^n(t)|\mu_j|\psi_i^n(t)\rangle\right] 
\end{equation}
where $j$ denotes the polarization direction of the electric field. The fidelity well adapted to take into
account the phases is given by:
\begin{equation}
 F=\frac{1}{Z^2} \lvert \sum_n^Z \langle \psi_i^n(t) | \phi_f^n \rangle \lvert^2
\end{equation}

\subsection{Deutsch-Josza algorithm}

The Deutsch-Josza algorithm for a one-qubit
function will serve us to illustrate the realization of phase correct gates

\begin{figure}
\includegraphics[width=0.7\columnwidth]{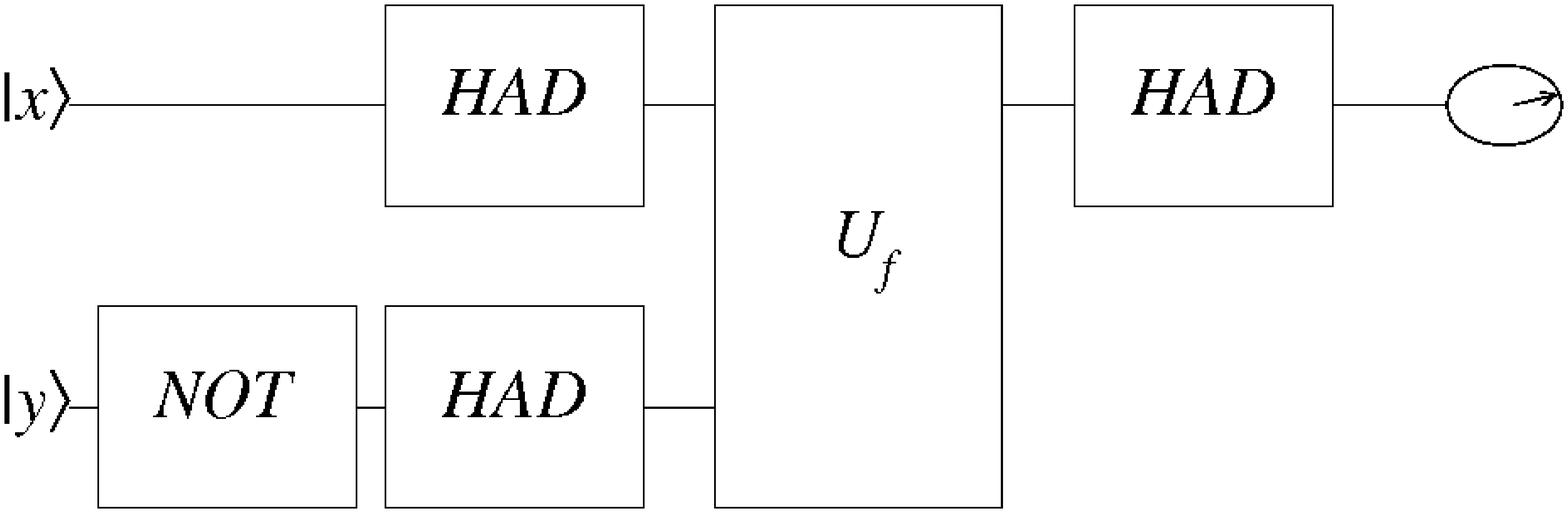}
\caption{Logical circuit for the Deutsch-Jozsa algorithm}
\label{Deutsch-Jozsa-logical-circuits}
\end{figure}

The Deutsch-Josza's problem illustrates the speedup of quantum computing by taking advantages of superposed
states. It has been the subject of several theoretical studies
\cite{Tesch2004,Ohtsuki2005,Ndong2007,Shioya2007,Mishima2009,Mishima2010}, and has been experimentally
implemented \cite{Takeushi2000,Dorai2000,Vala2002,Ju2010}.
The principles can be summarized as follows: we suppose that a function applied on one qubit $|x\rangle$ can either
change its value (balanced function) or leave it unchanged (constant function). The problem is to determine
whether the function is balanced or constant by a single call to the function and one measure. The logical circuit for the Deutsch-Jozsa algorithm is sketched Fig. 
\ref{Deutsch-Jozsa-logical-circuits}. 
The algorithm requires an auxiliary qubit $|y\rangle$. After a NOT gate on $|y\rangle$, the two qubits are put in a superposed state by one Hadamard gate on each qubit, The latter superposes 
the qubit states according to 
\begin{eqnarray}
 |0\rangle&\xrightarrow{HAD}&(1/\sqrt{2})(|0\rangle+|1\rangle)\\
|1\rangle&\xrightarrow{HAD}&(1/\sqrt{2})(|0\rangle-|1\rangle)
\end{eqnarray}
The call function is then implemented by applying the transformation $U_f$: 
\begin{equation}
U_f:|x\rangle|y\rangle\rightarrow |x\rangle|y\oplus f(x)\rangle
\end{equation}
 A final Hadamard gate is carried out on $|x\rangle$. 
When the initial state is $|00\rangle$, the final state of $|x\rangle$ is $\pm |0\rangle$ for a constant function, and $\pm |1\rangle$ for a balanced function. 
The nature of the
function is therefore determined by only one query. 

We simulate the example with the balanced function $f(0) = 1$ and $f(1) = 0$. In this case $U_f$ is the CNOT gate with control qubit $|x\rangle$. Qubits $|x\rangle$ and $|y\rangle$ are encoded in the first two
rotational states of two neighbouring molecules in the vibrational state $v = 0$: 
$|x\rangle|y\rangle\leftrightarrow|\tilde{N}_1\tilde{N}_2\rangle$ with $v_1=v_2=0$. (see Fig. \ref{Stark_Scheme}). The basis set for the simulation contains the vibrational 
state $v=0$ and rotational states up to $N=4$ for each molecule. Both Hadamard steps involve intramolecular transitions and the frequencies are mainly fixed by the Stark levels so
that the pulse duration is shorter than for the intermolecular CNOT gate which is the bottleneck of the full algorithm.

\subsection{NOT-HADHAD step}

\begin{figure}
\includegraphics[width=0.9\columnwidth]{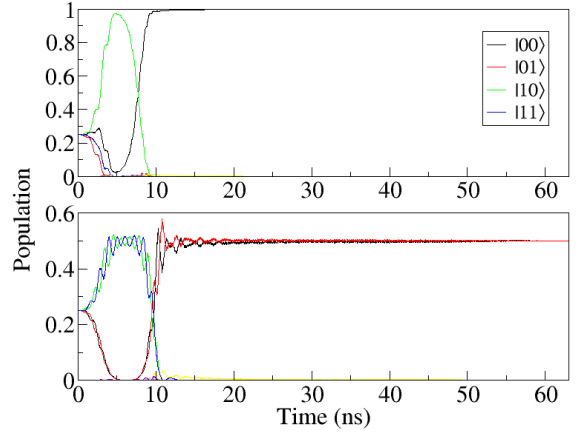}
\caption{Evolution of the populations during the gates involving HADAMARD gates in the Deutsch-Josza algorithm. Top panel: NOT-HADHAD step (NOT gate on $|y\rangle$ and the two 
HAD gates on $|x\rangle$ and $|y\rangle$. Bottom panel: HAD gate on $|x\rangle$. The initial state is a superposed state with equal weights 0.25 on the computational basis set 
states $|\tilde{0}\tilde{0}\rangle$, $|\tilde{0}\tilde{1}\rangle$, $|\tilde{1}\tilde{0}\rangle$, $|\tilde{1}\tilde{1}\rangle$.}
\label{HAD}
\end{figure}

\begin{figure}
\includegraphics[width=0.9\columnwidth]{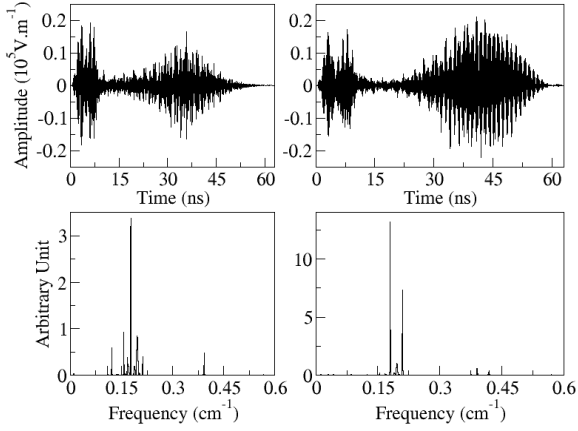}
\caption{Amplitude of the optimal field for the NOT-HADHAD gate (left top panel) and for the HAD gate (right top panel). Bottom panel; the corresponding Fourier transforms.}
\label{FIELDHH}
\end{figure}

 Optimal control allows us to optimize a global gate driving the resulting transformation of the first NOT 
gate on $|y\rangle$ and the two HAD gates on both qubits $|x\rangle$ and $|y\rangle$ in a single shot as suggested in previous works \cite{Weidinger2007, Bomble2008b, Bomble2009, Bomble2010}. 
This provides an interesting speedup. This NOT-HADHAD transformation in the coupled basis set is 
\begin{eqnarray}
|\tilde{0}\tilde{0}>&\rightarrow&2^{-1}(|\tilde{0}\tilde{0}\rangle-|\tilde{0}\tilde{1}\rangle+|\tilde{1}\tilde{0}\rangle-|\tilde{1}\tilde{1}\rangle)\\
|\tilde{0}\tilde{1}>&\rightarrow&2^{-1}(|\tilde{0}\tilde{0}\rangle+|\tilde{0}\tilde{1}\rangle+|\tilde{1}\tilde{0}\rangle+|\tilde{1}\tilde{1}\rangle)\\
|\tilde{1}\tilde{0}>&\rightarrow&2^{-1}(|\tilde{0}\tilde{0}\rangle+|\tilde{0}\tilde{1}\rangle-|\tilde{1}\tilde{0}\rangle-|\tilde{1}\tilde{1}\rangle)\\
|\tilde{1}\tilde{1}>&\rightarrow&2^{-1}(|\tilde{0}\tilde{0}\rangle-|\tilde{0}\tilde{1}\rangle-|\tilde{1}\tilde{0}\rangle+|\tilde{1}\tilde{1}\rangle)
\end{eqnarray}
and the additional phase equation is
\begin{equation}
2^{-1}(|\tilde{0}\tilde{0}\rangle+|\tilde{0}\tilde{1}\rangle)+|\tilde{1}\tilde{0}\rangle+|\tilde{1}\tilde{1}\rangle)\rightarrow |\tilde{0}\tilde{0}\rangle
\end{equation}
The NOT-HADHAD gate has been implemented by optimal control without guess field and a pulse duration $\tau_p = 63$ ns. A fidelity of 0.99999 is reached after 400 iterations. The
phases are optimized within $3\times10^{-3} \pi$. The evolution of the populations starting from a superposed state with equal weights 0.25 on the computational basis set states
$|\tilde{0}\tilde{0}\rangle$, $|\tilde{0}\tilde{1}\rangle$, $|\tilde{1}\tilde{0}\rangle$, $|\tilde{1}\tilde{1}\rangle$ is shown in top panel of Fig \ref{HAD}. The corresponding optimal field and its Fourier
transform are given in Fig.(\ref{FIELDHH}). 

\subsection{Phase correct intermolecular CNOT gate}

\begin{figure}
\includegraphics[width=0.9\columnwidth]{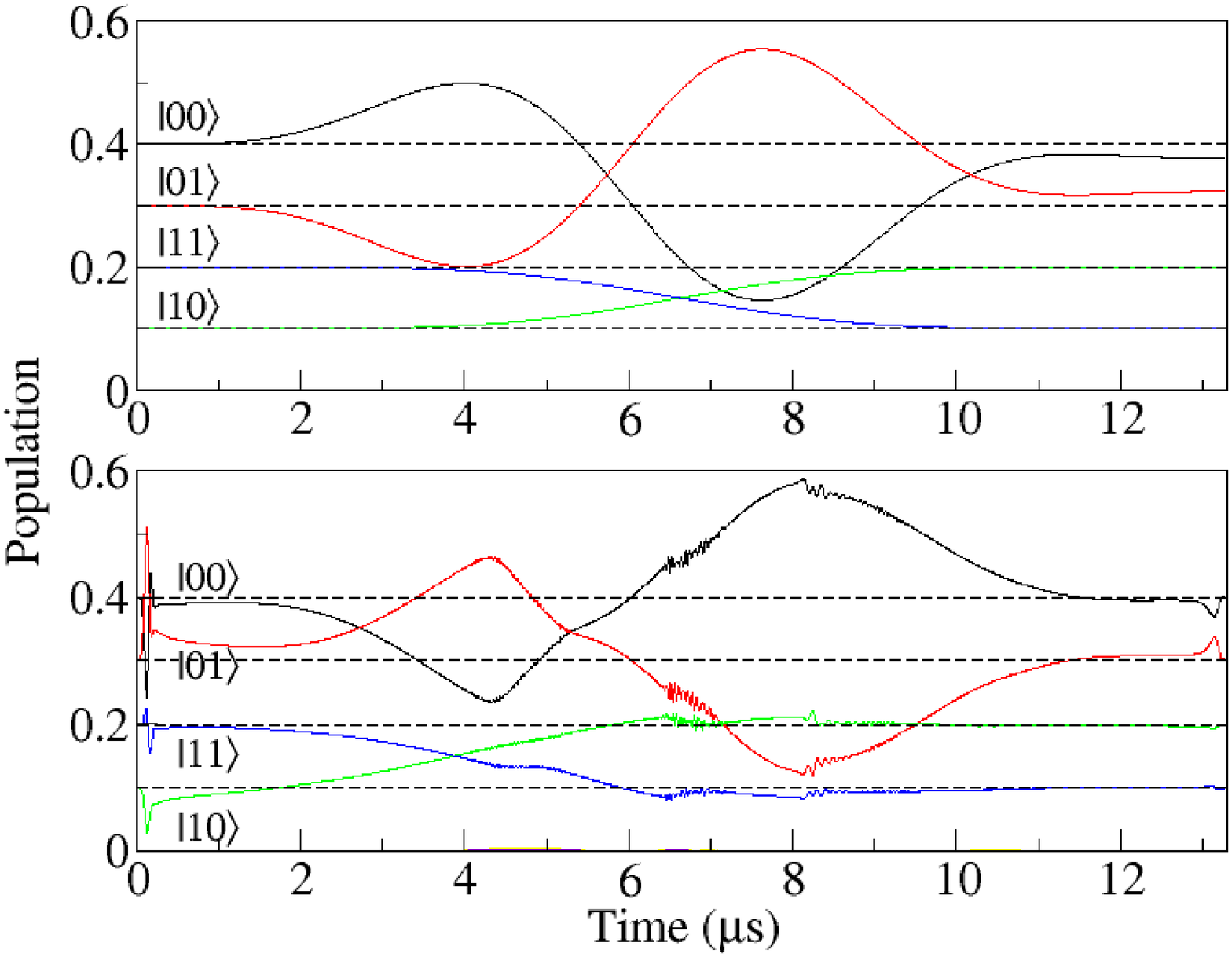}
\caption{Evolution of the populations during the phase CNOT gate starting from a superposition of
states $|\tilde{0}\tilde{0}>$, $|\tilde{0}\tilde{1}>$,$|\tilde{1}\tilde{0}>$,
$|\tilde{1}\tilde{1}>$. Top panel: gate driven by the $\pi$-pulse calculated in Fig. \ref{CNOTPI}. The pulse optimized for a pure state does not lead to a high fidelity operation when applied to a superposition of states; bottom panel: high fidelity gate optimized by optimal control. We checked that high fidelity gates were obtained when the pulse were applied to other initial superpositions. }
\label{CNOT_PHASE}
\end{figure}

 The supplementary transformation to ensure that the
phases of the final states of each transition are equal is here
\begin{multline}
\label{FithTransformation}
\frac{1}{2}(|00\rangle+|01\rangle+|10\rangle+|11\rangle)\rightarrow \\ 
\frac{1}{2}(|00\rangle e^{i\varphi_1}+|01\rangle e^{i\varphi_2} +|10\rangle e^{i\varphi_3}+|11\rangle e^{i\varphi_4})
\end{multline}
with
\begin{equation}
 \varphi_1=\varphi_2=\varphi_3=\varphi_4=\varphi
\end{equation}

where the phase $\varphi$ can take any value between 0 and $2\pi$. The $\pi$ pulse is a very good trial field since the convergence is fast with a fidelity index of
0.99943 in 130 iterations. Fig.\ref{CNOT_PHASE} compares the population evolution starting from a superposed state driven by the $\pi$ pulse 
(see Fig.\ref{CNOTPI}) or by the optimal control field. The fidelity for the transformations of the states $|\tilde{0}\tilde{0}\rangle$, $|\tilde{0}\tilde{1}\rangle$ decreases with the $\pi$
pulse. Owing to the phase constraint, the fidelity is better with optimal control. The phases obtained for the fifth transformation (\ref{FithTransformation}) are $\varphi_1 =
0.202 \pi$, $\varphi_2 = 0.199 \pi$, $\varphi_3 = 0.200 \pi$, and $\varphi_4 = 0.202 \pi$. The optimal control has found field amplitudes of the same order of magnitude
as for the $\pi$-pulse. The main frequency is the carrier frequency of the $\pi$ pulse and a lot of very small frequencies without direct signification from the model.

.

 \begin{figure}
\includegraphics[width=0.9\columnwidth]{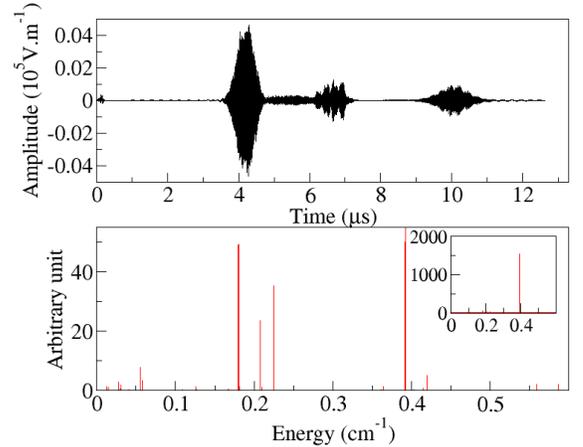}
\caption{Amplitude of the optimal field for the CNOT gate (top panel) and its Fourier transform (bottom panel).}
\label{FIELDCNOT}
\end{figure}
 
\subsection{HADAMARD gate on $|x\rangle$}

  One has to consider the extended transformation in a two-qubit space here. The HAD gate
(last step) on $|x\rangle$ becomes in the coupled basis set:
\begin{eqnarray}
|\tilde{0}\tilde{0}\rangle&\rightarrow&2^{-1/2}(|\tilde{0}\tilde{0}\rangle+|\tilde{0}\tilde{1}\rangle)\\
|\tilde{0}\tilde{1}\rangle&\rightarrow&2^{-1/2}(|\tilde{0}\tilde{1}\rangle+|\tilde{1}\tilde{1}\rangle)\\
|\tilde{1}\tilde{0}\rangle&\rightarrow&2^{-1/2}(|\tilde{0}\tilde{0}\rangle-|\tilde{1}\tilde{0}\rangle)\\
|\tilde{1}\tilde{1}\rangle&\rightarrow&2^{-1/2}(|\tilde{0}\tilde{1}\rangle-|\tilde{1}\tilde{1}\rangle)
\end{eqnarray}
with the additional equation for the phase constraint.
\begin{equation}
(|\tilde{0}\tilde{0}\rangle+|\tilde{0}\tilde{1}\rangle+|\tilde{1}\tilde{0}\rangle+|\tilde{1}\tilde{1}\rangle)\rightarrow
2^{-1/2}(|\tilde{0}\tilde{0}\rangle+|\tilde{0}\tilde{1}\rangle)
\end{equation}
The evolution of the populations starting from a superposed state with equal weights 0.25 on the computational basis set states $|\tilde{0}\tilde{0}\rangle$,
$|\tilde{0}\tilde{1}\rangle$, $|\tilde{1}\tilde{0}\rangle$, $|\tilde{1}\tilde{1}\rangle$ is shown in bottom panel of Fig \ref{HAD}. The corresponding optimal field and its Fourier
transform are given in Fig.\ref{FIELDHH}. As before, a fidelity of 0.99998 is obtained after 400 iterations. The phases are optimized within $10^{-2}\pi$.

Final optimal fields driving the desired operations can have a complicated envelop and differ drastically from the initial $\pi$-pulse raising the question of the robutness and actual realisation of such pulses. 
For schemes involving only rotational levels, the Fourier transform of the pulses show that the spectrum  contains frequencies in the microwave region only. In the case of the
HADAMARD gates (see Fig.\ref{FIELDHH}), 
frequencies are within 0.1 cm$^{-1}$ (3 GHz) and 0.2 cm$^{-1}$ (6 GHz)).  For this range of frequencies, electronic equipments can generate any arbitrary pulses and therefore obtaining such optimized pulses is experimentally feasible. 
Because several rotational levels are taken into account in the model (up to $N$=4 in the present study), optimized pulses may contain frequencies that will be more challenging to
obtain. In the case of the optimized CNOT gate (see Fig.\ref{FIELDCNOT}), the pulse contains frequencies up to 0.75 cm$^{-1}$   (20GHz). Filters can be used during the optimization procedure to eliminate out of range frequencies. 
If different vibrational levels are used. Pulses will be in the infrared region where pulse shaping techniques are less developed. 

\section{Conclusion}

Manipulating trapped ultracold molecules by laser fields offers great potentialities toward scalable
quantum computing. Logical operations can be performed by splitting processes into intramolecular global gates and intermolecular gates. The latter are very crucial for scalable operations for they enable the transfer of informations from one molecule to another through the dipole-dipole interaction. 
Using the states of the coupled basis set as logical states complicates the realization of one-qubit or two-qubit gates when the number of qubits involved in the logical operations increases. 
This can be overcome by using schemes that involve switchable interactions.
 Storing molecules in rovibrational levels with vanishing average dipole moment and transfering them when needed to levels with strong dipole moments by
adiabatic passage techniques could 
be a possibility. This operation would not take any longer than a few microsecond and would be equivalent in terms of duration to an additional intermolecular gate. 
The implementation of schemes that would include switchable interactions would further extend the potentialities of polar molecules for scalable quantum information. 
This remains open and can stimulate new researchs in molecular design, storage and
communications among entities.  \\

\section{Acknowledgments}
The authors wish to thank J. M. Teuler for technical assistance.
This research was supported by Triangle de la Physique under contract 20009-038T. The Computing facilities with the financial support of the FRS for the University of Li\`ege "Nic3" project is also
acknowledged. 

\bibliography{QuantumComputing.bib}

\begin{thebibliography}{58}
\expandafter\ifx\csname natexlab\endcsname\relax\def\natexlab#1{#1}\fi
\expandafter\ifx\csname bibnamefont\endcsname\relax
  \def\bibnamefont#1{#1}\fi
\expandafter\ifx\csname bibfnamefont\endcsname\relax
  \def\bibfnamefont#1{#1}\fi
\expandafter\ifx\csname citenamefont\endcsname\relax
  \def\citenamefont#1{#1}\fi
\expandafter\ifx\csname url\endcsname\relax
  \def\url#1{\texttt{#1}}\fi
\expandafter\ifx\csname urlprefix\endcsname\relax\def\urlprefix{URL }\fi
\providecommand{\bibinfo}[2]{#2}
\providecommand{\eprint}[2][]{\url{#2}}

\bibitem[{\citenamefont{Isenhower et~al.}(2010)\citenamefont{Isenhower, Urban,
  Zhang, Gill, Henage, Johnson, Walker, and Saffman}}]{Isenhower2010}
\bibinfo{author}{\bibfnamefont{L.}~\bibnamefont{Isenhower}},
  \bibinfo{author}{\bibfnamefont{E.}~\bibnamefont{Urban}},
  \bibinfo{author}{\bibfnamefont{X.~L.} \bibnamefont{Zhang}},
  \bibinfo{author}{\bibfnamefont{A.~T.} \bibnamefont{Gill}},
  \bibinfo{author}{\bibfnamefont{T.}~\bibnamefont{Henage}},
  \bibinfo{author}{\bibfnamefont{T.~A.} \bibnamefont{Johnson}},
  \bibinfo{author}{\bibfnamefont{T.~G.} \bibnamefont{Walker}},
  \bibnamefont{and} \bibinfo{author}{\bibfnamefont{M.}~\bibnamefont{Saffman}},
  \bibinfo{journal}{Phys. Rev. Lett.} \textbf{\bibinfo{volume}{104}},
  \bibinfo{pages}{010503} (\bibinfo{year}{2010}).

\bibitem[{\citenamefont{Monz et~al.}(2009)\citenamefont{Monz, Kim, Hänsel,
  Riebe, Villar, Schindler, Chwalla, Hennrich, and Blatt}}]{Monz2009}
\bibinfo{author}{\bibfnamefont{T.}~\bibnamefont{Monz}},
  \bibinfo{author}{\bibfnamefont{K.}~\bibnamefont{Kim}},
  \bibinfo{author}{\bibfnamefont{W.}~\bibnamefont{Hänsel}},
  \bibinfo{author}{\bibfnamefont{M.}~\bibnamefont{Riebe}},
  \bibinfo{author}{\bibfnamefont{A.~S.} \bibnamefont{Villar}},
  \bibinfo{author}{\bibfnamefont{P.}~\bibnamefont{Schindler}},
  \bibinfo{author}{\bibfnamefont{M.}~\bibnamefont{Chwalla}},
  \bibinfo{author}{\bibfnamefont{M.}~\bibnamefont{Hennrich}}, \bibnamefont{and}
  \bibinfo{author}{\bibfnamefont{R.}~\bibnamefont{Blatt}},
  \bibinfo{journal}{Phys. Rev. Lett.} \textbf{\bibinfo{volume}{102}},
  \bibinfo{pages}{040501} (\bibinfo{year}{2009}).

\bibitem[{\citenamefont{DiCarlo et~al.}(2009)\citenamefont{DiCarlo, Chow,
  Gambetta, Bishop, Johnson, Schuster, Majer, Blais, Frunzio, Girvin
  et~al.}}]{DiCarlo2009}
\bibinfo{author}{\bibfnamefont{L.}~\bibnamefont{DiCarlo}},
  \bibinfo{author}{\bibfnamefont{J.~M.} \bibnamefont{Chow}},
  \bibinfo{author}{\bibfnamefont{J.~M.} \bibnamefont{Gambetta}},
  \bibinfo{author}{\bibfnamefont{L.~S.} \bibnamefont{Bishop}},
  \bibinfo{author}{\bibfnamefont{B.~R.} \bibnamefont{Johnson}},
  \bibinfo{author}{\bibfnamefont{D.~I.} \bibnamefont{Schuster}},
  \bibinfo{author}{\bibfnamefont{J.}~\bibnamefont{Majer}},
  \bibinfo{author}{\bibfnamefont{A.}~\bibnamefont{Blais}},
  \bibinfo{author}{\bibfnamefont{L.}~\bibnamefont{Frunzio}},
  \bibinfo{author}{\bibfnamefont{S.~M.} \bibnamefont{Girvin}},
  \bibnamefont{et~al.}, \bibinfo{journal}{Nature}
  \textbf{\bibinfo{volume}{460}}, \bibinfo{pages}{240} (\bibinfo{year}{2009}).

\bibitem[{\citenamefont{Politi et~al.}(2009)\citenamefont{Politi, Matthews, and
  O'Brien}}]{Politi2009}
\bibinfo{author}{\bibfnamefont{A.}~\bibnamefont{Politi}},
  \bibinfo{author}{\bibfnamefont{J.~C.~F.} \bibnamefont{Matthews}},
  \bibnamefont{and} \bibinfo{author}{\bibfnamefont{J.~L.}
  \bibnamefont{O'Brien}}, \bibinfo{journal}{Science}
  \textbf{\bibinfo{volume}{325}}, \bibinfo{pages}{1221} (\bibinfo{year}{2009}).

\bibitem[{\citenamefont{Du et~al.}(2010)\citenamefont{Du, Xu, Peng, Wang, Wu,
  and Lu}}]{Du2010}
\bibinfo{author}{\bibfnamefont{J.}~\bibnamefont{Du}},
  \bibinfo{author}{\bibfnamefont{N.}~\bibnamefont{Xu}},
  \bibinfo{author}{\bibfnamefont{X.}~\bibnamefont{Peng}},
  \bibinfo{author}{\bibfnamefont{P.}~\bibnamefont{Wang}},
  \bibinfo{author}{\bibfnamefont{S.}~\bibnamefont{Wu}}, \bibnamefont{and}
  \bibinfo{author}{\bibfnamefont{D.}~\bibnamefont{Lu}}, \bibinfo{journal}{Phys.
  Rev. Lett.} \textbf{\bibinfo{volume}{104}}, \bibinfo{pages}{030502}
  (\bibinfo{year}{2010}).

\bibitem[{\citenamefont{Tesch et~al.}(2001)\citenamefont{Tesch, Kurtz, and
  de~Vivie-Riedle}}]{Tesch2001}
\bibinfo{author}{\bibfnamefont{C.~M.} \bibnamefont{Tesch}},
  \bibinfo{author}{\bibfnamefont{L.}~\bibnamefont{Kurtz}}, \bibnamefont{and}
  \bibinfo{author}{\bibfnamefont{R.}~\bibnamefont{de~Vivie-Riedle}},
  \bibinfo{journal}{Chem. Phys. Lett.} \textbf{\bibinfo{volume}{343}},
  \bibinfo{pages}{633} (\bibinfo{year}{2001}).

\bibitem[{\citenamefont{Tesch and de~Vivie-Riedle}(2002)}]{Tesch2002}
\bibinfo{author}{\bibfnamefont{C.~M.} \bibnamefont{Tesch}} \bibnamefont{and}
  \bibinfo{author}{\bibfnamefont{R.}~\bibnamefont{de~Vivie-Riedle}},
  \bibinfo{journal}{Phys. Rev. Lett} \textbf{\bibinfo{volume}{89}},
  \bibinfo{pages}{157901} (\bibinfo{year}{2002}).

\bibitem[{\citenamefont{Vala1 et~al.}(2002)\citenamefont{Vala1, Amitay, Zhang,
  Leone, and Kosloff}}]{Vala2002}
\bibinfo{author}{\bibfnamefont{J.}~\bibnamefont{Vala1}},
  \bibinfo{author}{\bibfnamefont{Z.}~\bibnamefont{Amitay}},
  \bibinfo{author}{\bibfnamefont{B.}~\bibnamefont{Zhang}},
  \bibinfo{author}{\bibfnamefont{S.~R.} \bibnamefont{Leone}}, \bibnamefont{and}
  \bibinfo{author}{\bibfnamefont{R.}~\bibnamefont{Kosloff}},
  \bibinfo{journal}{Phys. Rev. A} \textbf{\bibinfo{volume}{66}},
  \bibinfo{pages}{062316} (\bibinfo{year}{2002}).

\bibitem[{\citenamefont{Babikov}(2004)}]{Babikov2004}
\bibinfo{author}{\bibfnamefont{D.}~\bibnamefont{Babikov}}, \bibinfo{journal}{J.
  Chem. Phys.} \textbf{\bibinfo{volume}{121}}, \bibinfo{pages}{7577}
  (\bibinfo{year}{2004}).

\bibitem[{\citenamefont{Ohtsuki}(2005)}]{Ohtsuki2005}
\bibinfo{author}{\bibfnamefont{Y.}~\bibnamefont{Ohtsuki}},
  \bibinfo{journal}{Chem. Phys. Lett.} \textbf{\bibinfo{volume}{404}},
  \bibinfo{pages}{126} (\bibinfo{year}{2005}).

\bibitem[{\citenamefont{Menzel-Jones and Shapiro}(2007)}]{Menzel2007}
\bibinfo{author}{\bibfnamefont{C.}~\bibnamefont{Menzel-Jones}}
  \bibnamefont{and} \bibinfo{author}{\bibfnamefont{M.}~\bibnamefont{Shapiro}},
  \bibinfo{journal}{Phys. Rev. A} \textbf{\bibinfo{volume}{75}},
  \bibinfo{pages}{052308} (\bibinfo{year}{2007}).

\bibitem[{\citenamefont{Shioya et~al.}(2007)\citenamefont{Shioya, Mishima, and
  Yamashita}}]{Shioya2007}
\bibinfo{author}{\bibfnamefont{K.}~\bibnamefont{Shioya}},
  \bibinfo{author}{\bibfnamefont{K.}~\bibnamefont{Mishima}}, \bibnamefont{and}
  \bibinfo{author}{\bibfnamefont{K.}~\bibnamefont{Yamashita}},
  \bibinfo{journal}{Mol. Phys.} \textbf{\bibinfo{volume}{105}},
  \bibinfo{pages}{1287} (\bibinfo{year}{2007}).

\bibitem[{\citenamefont{Mishima et~al.}(2008)\citenamefont{Mishima, Tokumo, and
  Yamashita}}]{Mishima2008}
\bibinfo{author}{\bibfnamefont{K.}~\bibnamefont{Mishima}},
  \bibinfo{author}{\bibfnamefont{K.}~\bibnamefont{Tokumo}}, \bibnamefont{and}
  \bibinfo{author}{\bibfnamefont{K.}~\bibnamefont{Yamashita}},
  \bibinfo{journal}{Chem. Phys.} \textbf{\bibinfo{volume}{343}},
  \bibinfo{pages}{61} (\bibinfo{year}{2008}).

\bibitem[{\citenamefont{Tsubouchi and Momose}(2008)}]{Tsubouchi2008}
\bibinfo{author}{\bibfnamefont{M.}~\bibnamefont{Tsubouchi}} \bibnamefont{and}
  \bibinfo{author}{\bibfnamefont{T.}~\bibnamefont{Momose}},
  \bibinfo{journal}{Phys. Rev. A} \textbf{\bibinfo{volume}{77}},
  \bibinfo{pages}{052326} (\bibinfo{year}{2008}).

\bibitem[{\citenamefont{Tsubouchi et~al.}(2008)\citenamefont{Tsubouchi,
  Khramov, and Momose}}]{Tsubouchi2008b}
\bibinfo{author}{\bibfnamefont{M.}~\bibnamefont{Tsubouchi}},
  \bibinfo{author}{\bibfnamefont{A.}~\bibnamefont{Khramov}}, \bibnamefont{and}
  \bibinfo{author}{\bibfnamefont{T.}~\bibnamefont{Momose}},
  \bibinfo{journal}{Phys. Rev. A} \textbf{\bibinfo{volume}{77}},
  \bibinfo{pages}{023405} (\bibinfo{year}{2008}).

\bibitem[{\citenamefont{Sugny et~al.}(2009)\citenamefont{Sugny, Bomble,
  Ribeyre, Dulieu, and Desouter-Lecomte}}]{Sugny2009}
\bibinfo{author}{\bibfnamefont{D.}~\bibnamefont{Sugny}},
  \bibinfo{author}{\bibfnamefont{L.}~\bibnamefont{Bomble}},
  \bibinfo{author}{\bibfnamefont{T.}~\bibnamefont{Ribeyre}},
  \bibinfo{author}{\bibfnamefont{O.}~\bibnamefont{Dulieu}}, \bibnamefont{and}
  \bibinfo{author}{\bibfnamefont{M.}~\bibnamefont{Desouter-Lecomte}},
  \bibinfo{journal}{Phys. Rev. A} \textbf{\bibinfo{volume}{80}},
  \bibinfo{pages}{042325} (\bibinfo{year}{2009}).

\bibitem[{\citenamefont{Mishima and Yamashita}(2010)}]{Mishima2010}
\bibinfo{author}{\bibfnamefont{K.}~\bibnamefont{Mishima}} \bibnamefont{and}
  \bibinfo{author}{\bibfnamefont{K.}~\bibnamefont{Yamashita}},
  \bibinfo{journal}{Chem. Phys.} \textbf{\bibinfo{volume}{376}},
  \bibinfo{pages}{63} (\bibinfo{year}{2010}).

\bibitem[{\citenamefont{Zaari and Brown}(2010)}]{Zaari2010}
\bibinfo{author}{\bibfnamefont{R.}~\bibnamefont{Zaari}} \bibnamefont{and}
  \bibinfo{author}{\bibfnamefont{A.}~\bibnamefont{Brown}}, \bibinfo{journal}{J.
  Chem. Phys.} \textbf{\bibinfo{volume}{132}}, \bibinfo{pages}{014307}
  (\bibinfo{year}{2010}).

\bibitem[{\citenamefont{Tesch and de~Vivie-Riedle}(2004)}]{Tesch2004}
\bibinfo{author}{\bibfnamefont{C.~M.} \bibnamefont{Tesch}} \bibnamefont{and}
  \bibinfo{author}{\bibfnamefont{R.}~\bibnamefont{de~Vivie-Riedle}},
  \bibinfo{journal}{J. Chem. Phys.} \textbf{\bibinfo{volume}{121}},
  \bibinfo{pages}{12158} (\bibinfo{year}{2004}).

\bibitem[{\citenamefont{Troppmann and de~Vivie-Riedle}(2005)}]{Troppmann2005}
\bibinfo{author}{\bibfnamefont{U.}~\bibnamefont{Troppmann}} \bibnamefont{and}
  \bibinfo{author}{\bibfnamefont{R.}~\bibnamefont{de~Vivie-Riedle}},
  \bibinfo{journal}{J. Chem. Phys.} \textbf{\bibinfo{volume}{122}},
  \bibinfo{pages}{154105} (\bibinfo{year}{2005}).

\bibitem[{\citenamefont{Korff et~al.}(2005)\citenamefont{Korff, Troppmann,
  Kompa, and de~Vivie-Riedle}}]{Korff2005}
\bibinfo{author}{\bibfnamefont{B.}~\bibnamefont{Korff}},
  \bibinfo{author}{\bibfnamefont{U.}~\bibnamefont{Troppmann}},
  \bibinfo{author}{\bibfnamefont{K.}~\bibnamefont{Kompa}}, \bibnamefont{and}
  \bibinfo{author}{\bibfnamefont{R.}~\bibnamefont{de~Vivie-Riedle}},
  \bibinfo{journal}{J. Chem. Phys.} \textbf{\bibinfo{volume}{123}},
  \bibinfo{pages}{244509} (\bibinfo{year}{2005}).

\bibitem[{\citenamefont{Ndong et~al.}(2006)\citenamefont{Ndong, Bomble, Sugny,
  Justum, and Desouter-Lecomte}}]{Sugny2006}
\bibinfo{author}{\bibfnamefont{M.}~\bibnamefont{Ndong}},
  \bibinfo{author}{\bibfnamefont{L.}~\bibnamefont{Bomble}},
  \bibinfo{author}{\bibfnamefont{D.}~\bibnamefont{Sugny}},
  \bibinfo{author}{\bibfnamefont{Y.}~\bibnamefont{Justum}}, \bibnamefont{and}
  \bibinfo{author}{\bibfnamefont{M.}~\bibnamefont{Desouter-Lecomte}},
  \bibinfo{journal}{Phys. Rev. A} \textbf{\bibinfo{volume}{74}},
  \bibinfo{pages}{043419} (\bibinfo{year}{2006}).

\bibitem[{\citenamefont{Sugny et~al.}(2007)\citenamefont{Sugny, Ndong,
  lauvergnat, Justum, and Desouter-Lecomte}}]{Sugny2007}
\bibinfo{author}{\bibfnamefont{D.}~\bibnamefont{Sugny}},
  \bibinfo{author}{\bibfnamefont{M.}~\bibnamefont{Ndong}},
  \bibinfo{author}{\bibfnamefont{D.}~\bibnamefont{lauvergnat}},
  \bibinfo{author}{\bibfnamefont{Y.}~\bibnamefont{Justum}}, \bibnamefont{and}
  \bibinfo{author}{\bibfnamefont{M.}~\bibnamefont{Desouter-Lecomte}},
  \bibinfo{journal}{J. Photochem. Photobiol A.} \textbf{\bibinfo{volume}{190}},
  \bibinfo{pages}{350} (\bibinfo{year}{2007}).

\bibitem[{\citenamefont{Ndong et~al.}(2007)\citenamefont{Ndong, Lauvergnat,
  Chapuisat, and Desouter-Lecomte}}]{Ndong2007}
\bibinfo{author}{\bibfnamefont{M.}~\bibnamefont{Ndong}},
  \bibinfo{author}{\bibfnamefont{D.}~\bibnamefont{Lauvergnat}},
  \bibinfo{author}{\bibfnamefont{X.}~\bibnamefont{Chapuisat}},
  \bibnamefont{and}
  \bibinfo{author}{\bibfnamefont{M.}~\bibnamefont{Desouter-Lecomte}},
  \bibinfo{journal}{J. Chem. Phys.} \textbf{\bibinfo{volume}{126}},
  \bibinfo{pages}{244505} (\bibinfo{year}{2007}).

\bibitem[{\citenamefont{Weidinger and Gruebele}(2007)}]{Weidinger2007}
\bibinfo{author}{\bibfnamefont{D.}~\bibnamefont{Weidinger}} \bibnamefont{and}
  \bibinfo{author}{\bibfnamefont{M.}~\bibnamefont{Gruebele}},
  \bibinfo{journal}{Mol. Phys.} \textbf{\bibinfo{volume}{105}},
  \bibinfo{pages}{1999} (\bibinfo{year}{2007}).

\bibitem[{\citenamefont{Bomble et~al.}(2008)\citenamefont{Bomble, Lauvergnat,
  Remacle, and Desouter-Lecomte}}]{Bomble2008b}
\bibinfo{author}{\bibfnamefont{L.}~\bibnamefont{Bomble}},
  \bibinfo{author}{\bibfnamefont{D.}~\bibnamefont{Lauvergnat}},
  \bibinfo{author}{\bibfnamefont{F.}~\bibnamefont{Remacle}}, \bibnamefont{and}
  \bibinfo{author}{\bibfnamefont{M.}~\bibnamefont{Desouter-Lecomte}},
  \bibinfo{journal}{J. Chem. Phys.} \textbf{\bibinfo{volume}{128}},
  \bibinfo{pages}{064110} (\bibinfo{year}{2008}).

\bibitem[{\citenamefont{Bomble et~al.}(2009)\citenamefont{Bomble, Lauvergnat,
  Remacle, and Desouter-Lecomte}}]{Bomble2009}
\bibinfo{author}{\bibfnamefont{L.}~\bibnamefont{Bomble}},
  \bibinfo{author}{\bibfnamefont{D.}~\bibnamefont{Lauvergnat}},
  \bibinfo{author}{\bibfnamefont{F.}~\bibnamefont{Remacle}}, \bibnamefont{and}
  \bibinfo{author}{\bibfnamefont{M.}~\bibnamefont{Desouter-Lecomte}},
  \bibinfo{journal}{Phys. Rev. A} \textbf{\bibinfo{volume}{80}},
  \bibinfo{pages}{022332} (\bibinfo{year}{2009}).

\bibitem[{\citenamefont{Bomble et~al.}(2010)\citenamefont{Bomble, Lauvergnat,
  Remacle, and Desouter-Lecomte}}]{Bomble2010}
\bibinfo{author}{\bibfnamefont{L.}~\bibnamefont{Bomble}},
  \bibinfo{author}{\bibfnamefont{D.}~\bibnamefont{Lauvergnat}},
  \bibinfo{author}{\bibfnamefont{F.}~\bibnamefont{Remacle}}, \bibnamefont{and}
  \bibinfo{author}{\bibfnamefont{M.}~\bibnamefont{Desouter-Lecomte}},
  \bibinfo{journal}{Phys. Chem. Chem. Phys.} \textbf{\bibinfo{volume}{in
  press}} (\bibinfo{year}{2010}).

\bibitem[{\citenamefont{Schr\"oder and Brown}(2009)}]{Schroder2009}
\bibinfo{author}{\bibfnamefont{M.}~\bibnamefont{Schr\"oder}} \bibnamefont{and}
  \bibinfo{author}{\bibfnamefont{A.}~\bibnamefont{Brown}}, \bibinfo{journal}{J.
  Chem. Phys.} \textbf{\bibinfo{volume}{131}}, \bibinfo{pages}{034101}
  (\bibinfo{year}{2009}).

\bibitem[{\citenamefont{Mishima and
  Yamashita}(2009{\natexlab{a}})}]{Mishima2009}
\bibinfo{author}{\bibfnamefont{K.}~\bibnamefont{Mishima}} \bibnamefont{and}
  \bibinfo{author}{\bibfnamefont{K.}~\bibnamefont{Yamashita}},
  \bibinfo{journal}{Chem. Phys.} \textbf{\bibinfo{volume}{361}},
  \bibinfo{pages}{106} (\bibinfo{year}{2009}{\natexlab{a}}).

\bibitem[{\citenamefont{Mishima and
  Yamashita}(2009{\natexlab{b}})}]{Mishima2009b}
\bibinfo{author}{\bibfnamefont{K.}~\bibnamefont{Mishima}} \bibnamefont{and}
  \bibinfo{author}{\bibfnamefont{K.}~\bibnamefont{Yamashita}},
  \bibinfo{journal}{J. Chem. Phys.} \textbf{\bibinfo{volume}{130}},
  \bibinfo{pages}{034108} (\bibinfo{year}{2009}{\natexlab{b}}).

\bibitem[{\citenamefont{DeMille}(2002)}]{DeMille2002}
\bibinfo{author}{\bibfnamefont{D.}~\bibnamefont{DeMille}},
  \bibinfo{journal}{Phys. Rev. Lett.} \textbf{\bibinfo{volume}{88}},
  \bibinfo{pages}{067901} (\bibinfo{year}{2002}).

\bibitem[{\citenamefont{Carr et~al.}(2009)\citenamefont{Carr, DeMille, Krems,
  and Ye}}]{Carr2009}
\bibinfo{author}{\bibfnamefont{L.~D.} \bibnamefont{Carr}},
  \bibinfo{author}{\bibfnamefont{D.}~\bibnamefont{DeMille}},
  \bibinfo{author}{\bibfnamefont{R.~V.} \bibnamefont{Krems}}, \bibnamefont{and}
  \bibinfo{author}{\bibfnamefont{J.}~\bibnamefont{Ye}}, \bibinfo{journal}{New
  J. Phys.} \textbf{\bibinfo{volume}{11}}, \bibinfo{pages}{055049}
  (\bibinfo{year}{2009}).

\bibitem[{\citenamefont{Yelin et~al.}(2006)\citenamefont{Yelin, Kirby, and
  C\^ot\'e}}]{Yelin2006}
\bibinfo{author}{\bibfnamefont{S.~F.} \bibnamefont{Yelin}},
  \bibinfo{author}{\bibfnamefont{K.}~\bibnamefont{Kirby}}, \bibnamefont{and}
  \bibinfo{author}{\bibfnamefont{R.}~\bibnamefont{C\^ot\'e}},
  \bibinfo{journal}{Phys. Rev. A} \textbf{\bibinfo{volume}{74}},
  \bibinfo{pages}{050301(R)} (\bibinfo{year}{2006}).

\bibitem[{\citenamefont{Kuznetsova et~al.}(2008)\citenamefont{Kuznetsova,
  C\^ot\'e, Kirby, and Yelin}}]{Kuznetsova2008}
\bibinfo{author}{\bibfnamefont{E.}~\bibnamefont{Kuznetsova}},
  \bibinfo{author}{\bibfnamefont{R.}~\bibnamefont{C\^ot\'e}},
  \bibinfo{author}{\bibfnamefont{K.}~\bibnamefont{Kirby}}, \bibnamefont{and}
  \bibinfo{author}{\bibfnamefont{S.~F.} \bibnamefont{Yelin}},
  \bibinfo{journal}{Phys. Rev. A} \textbf{\bibinfo{volume}{78}},
  \bibinfo{pages}{012313} (\bibinfo{year}{2008}).

\bibitem[{\citenamefont{Charron et~al.}(2007)\citenamefont{Charron, Milman,
  Keller, and Atabek}}]{Charron2007}
\bibinfo{author}{\bibfnamefont{E.}~\bibnamefont{Charron}},
  \bibinfo{author}{\bibfnamefont{P.}~\bibnamefont{Milman}},
  \bibinfo{author}{\bibfnamefont{A.}~\bibnamefont{Keller}}, \bibnamefont{and}
  \bibinfo{author}{\bibfnamefont{O.}~\bibnamefont{Atabek}},
  \bibinfo{journal}{Phys. Rev. A} \textbf{\bibinfo{volume}{75}},
  \bibinfo{pages}{033414} (\bibinfo{year}{2007}).

\bibitem[{\citenamefont{Beckman et~al.}(1996)\citenamefont{Beckman, Chari,
  Devabhaktuni, and Preskill}}]{Beckman1996}
\bibinfo{author}{\bibfnamefont{D.}~\bibnamefont{Beckman}},
  \bibinfo{author}{\bibfnamefont{A.}~\bibnamefont{Chari}},
  \bibinfo{author}{\bibfnamefont{S.}~\bibnamefont{Devabhaktuni}},
  \bibnamefont{and} \bibinfo{author}{\bibfnamefont{J.}~\bibnamefont{Preskill}},
  \bibinfo{journal}{Phys. Rev. A} \textbf{\bibinfo{volume}{54}},
  \bibinfo{pages}{1034} (\bibinfo{year}{1996}).

\bibitem[{\citenamefont{Deutsch and Jozsa}(1992)}]{Deutsch1992}
\bibinfo{author}{\bibfnamefont{D.}~\bibnamefont{Deutsch}} \bibnamefont{and}
  \bibinfo{author}{\bibfnamefont{R.}~\bibnamefont{Jozsa}},
  \bibinfo{journal}{Proc. R. Soc. Lond. A} \textbf{\bibinfo{volume}{449}},
  \bibinfo{pages}{669} (\bibinfo{year}{1992}).

\bibitem[{\citenamefont{Aymar and Dulieu}(2005)}]{Aymar2005}
\bibinfo{author}{\bibfnamefont{M.}~\bibnamefont{Aymar}} \bibnamefont{and}
  \bibinfo{author}{\bibfnamefont{O.}~\bibnamefont{Dulieu}},
  \bibinfo{journal}{J. Chem. Phys.} \textbf{\bibinfo{volume}{122}},
  \bibinfo{pages}{204302} (\bibinfo{year}{2005}).

\bibitem[{\citenamefont{Haimberger et~al.}(2004)\citenamefont{Haimberger,
  Kleinert, Bhattacharya, and Bigelow}}]{Haimberger2004}
\bibinfo{author}{\bibfnamefont{C.}~\bibnamefont{Haimberger}},
  \bibinfo{author}{\bibfnamefont{J.}~\bibnamefont{Kleinert}},
  \bibinfo{author}{\bibfnamefont{M.}~\bibnamefont{Bhattacharya}},
  \bibnamefont{and} \bibinfo{author}{\bibfnamefont{N.~P.}
  \bibnamefont{Bigelow}}, \bibinfo{journal}{Phys. Rev. A}
  \textbf{\bibinfo{volume}{70}}, \bibinfo{pages}{021402}
  (\bibinfo{year}{2004}).

\bibitem[{\citenamefont{Korek et~al.}(2000)\citenamefont{Korek, Allouche,
  Fakhreddine, and Chaalan}}]{Korek2000}
\bibinfo{author}{\bibfnamefont{M.}~\bibnamefont{Korek}},
  \bibinfo{author}{\bibfnamefont{A.~R.} \bibnamefont{Allouche}},
  \bibinfo{author}{\bibfnamefont{K.}~\bibnamefont{Fakhreddine}},
  \bibnamefont{and} \bibinfo{author}{\bibfnamefont{A.}~\bibnamefont{Chaalan}},
  \bibinfo{journal}{Can. J. Phys.} \textbf{\bibinfo{volume}{78}},
  \bibinfo{pages}{977} (\bibinfo{year}{2000}).

\bibitem[{\citenamefont{Kotochigova and Tiesinga}(2006)}]{Kotochigova2006}
\bibinfo{author}{\bibfnamefont{S.}~\bibnamefont{Kotochigova}} \bibnamefont{and}
  \bibinfo{author}{\bibfnamefont{E.}~\bibnamefont{Tiesinga}},
  \bibinfo{journal}{Phys. Rev. A} \textbf{\bibinfo{volume}{73}},
  \bibinfo{pages}{041405(R)} (\bibinfo{year}{2006}).

\bibitem[{\citenamefont{Micheli et~al.}(2007)\citenamefont{Micheli, Pupillo,
  B\"uchler, and Zoller}}]{Micheli2007}
\bibinfo{author}{\bibfnamefont{A.}~\bibnamefont{Micheli}},
  \bibinfo{author}{\bibfnamefont{G.}~\bibnamefont{Pupillo}},
  \bibinfo{author}{\bibfnamefont{H.~P.} \bibnamefont{B\"uchler}},
  \bibnamefont{and} \bibinfo{author}{\bibfnamefont{P.}~\bibnamefont{Zoller}},
  \bibinfo{journal}{Phys. Rev. A} \textbf{\bibinfo{volume}{76}},
  \bibinfo{pages}{043604} (\bibinfo{year}{2007}).

\bibitem[{\citenamefont{Cohen-Tannoudji
  et~al.}(1973)\citenamefont{Cohen-Tannoudji, Diu, and
  Lalo\"e}}]{Cohen-Tannoudji}
\bibinfo{author}{\bibfnamefont{C.}~\bibnamefont{Cohen-Tannoudji}},
  \bibinfo{author}{\bibfnamefont{B.}~\bibnamefont{Diu}}, \bibnamefont{and}
  \bibinfo{author}{\bibfnamefont{F.}~\bibnamefont{Lalo\"e}},
  \emph{\bibinfo{title}{M\'ecanique quantique}} (\bibinfo{publisher}{Hermann},
  \bibinfo{year}{1973}).

\bibitem[{\citenamefont{Press et~al.}(1986)\citenamefont{Press, Teukolsky,
  Vetterling, and Flannery}}]{NumericalRecipes}
\bibinfo{author}{\bibfnamefont{W.~H.} \bibnamefont{Press}},
  \bibinfo{author}{\bibfnamefont{S.~A.} \bibnamefont{Teukolsky}},
  \bibinfo{author}{\bibfnamefont{W.~T.} \bibnamefont{Vetterling}},
  \bibnamefont{and} \bibinfo{author}{\bibfnamefont{B.~P.}
  \bibnamefont{Flannery}}, \emph{\bibinfo{title}{Numerical Recipes in FORTRAN,
  The Art of Scientific Computing, Second Edition}}
  (\bibinfo{publisher}{Cambridge University Press}, \bibinfo{year}{1986}).

\bibitem[{\citenamefont{Barenco et~al.}(1995)\citenamefont{Barenco, Deutsch,
  and Ekert}}]{Barenco1995}
\bibinfo{author}{\bibfnamefont{A.}~\bibnamefont{Barenco}},
  \bibinfo{author}{\bibfnamefont{D.}~\bibnamefont{Deutsch}}, \bibnamefont{and}
  \bibinfo{author}{\bibfnamefont{A.}~\bibnamefont{Ekert}},
  \bibinfo{journal}{Phys. Rev. Lett.} \textbf{\bibinfo{volume}{74}},
  \bibinfo{pages}{4081} (\bibinfo{year}{1995}).

\bibitem[{\citenamefont{Berman et~al.}(1997)\citenamefont{Berman, Campbell,
  Doolen, L\'opez, and Tsifrinovich}}]{Berman1997}
\bibinfo{author}{\bibfnamefont{G.~P.} \bibnamefont{Berman}},
  \bibinfo{author}{\bibfnamefont{D.~K.} \bibnamefont{Campbell}},
  \bibinfo{author}{\bibfnamefont{G.~D.} \bibnamefont{Doolen}},
  \bibinfo{author}{\bibfnamefont{G.~V.} \bibnamefont{L\'opez}},
  \bibnamefont{and} \bibinfo{author}{\bibfnamefont{V.~I.}
  \bibnamefont{Tsifrinovich}}, \bibinfo{journal}{Physica B}
  \textbf{\bibinfo{volume}{240}}, \bibinfo{pages}{61} (\bibinfo{year}{1997}).

\bibitem[{\citenamefont{Vasilev and Vitanov}(2004)}]{Vasilev2004}
\bibinfo{author}{\bibfnamefont{G.}~\bibnamefont{Vasilev}} \bibnamefont{and}
  \bibinfo{author}{\bibfnamefont{N.~V.} \bibnamefont{Vitanov}},
  \bibinfo{journal}{Phys. Rev. A} \textbf{\bibinfo{volume}{70}},
  \bibinfo{pages}{053407} (\bibinfo{year}{2004}).

\bibitem[{\citenamefont{Vedral et~al.}(1996)\citenamefont{Vedral, Barenco, and
  Ekert}}]{Vedral1996}
\bibinfo{author}{\bibfnamefont{V.}~\bibnamefont{Vedral}},
  \bibinfo{author}{\bibfnamefont{A.}~\bibnamefont{Barenco}}, \bibnamefont{and}
  \bibinfo{author}{\bibfnamefont{A.}~\bibnamefont{Ekert}},
  \bibinfo{journal}{Phys. Rev. A} \textbf{\bibinfo{volume}{54}},
  \bibinfo{pages}{147} (\bibinfo{year}{1996}).

\bibitem[{\citenamefont{Benenti et~al.}(2004)\citenamefont{Benenti, Casati, and
  Strini}}]{Benenti}
\bibinfo{author}{\bibfnamefont{G.}~\bibnamefont{Benenti}},
  \bibinfo{author}{\bibfnamefont{G.}~\bibnamefont{Casati}}, \bibnamefont{and}
  \bibinfo{author}{\bibfnamefont{G.}~\bibnamefont{Strini}},
  \emph{\bibinfo{title}{Principles of Quantum Computation and Information}}
  (\bibinfo{publisher}{World Scientific Publishing Co. Pte. LTD},
  \bibinfo{year}{2004}).

\bibitem[{\citenamefont{Aldegunde et~al.}(2009)\citenamefont{Aldegunde, Ran,
  and Hutson}}]{Aldegunde2009}
\bibinfo{author}{\bibfnamefont{J.}~\bibnamefont{Aldegunde}},
  \bibinfo{author}{\bibfnamefont{H.}~\bibnamefont{Ran}}, \bibnamefont{and}
  \bibinfo{author}{\bibfnamefont{J.~M.} \bibnamefont{Hutson}},
  \bibinfo{journal}{Phys. Rev. A} \textbf{\bibinfo{volume}{80}},
  \bibinfo{pages}{043410} (\bibinfo{year}{2009}).

\bibitem[{\citenamefont{Ran et~al.}(2010)\citenamefont{Ran, Aldegunde, and
  Hutson}}]{Ran2010}
\bibinfo{author}{\bibfnamefont{H.}~\bibnamefont{Ran}},
  \bibinfo{author}{\bibfnamefont{J.}~\bibnamefont{Aldegunde}},
  \bibnamefont{and} \bibinfo{author}{\bibfnamefont{J.~M.}
  \bibnamefont{Hutson}}, \bibinfo{journal}{New J. Phys.}
  \textbf{\bibinfo{volume}{12}}, \bibinfo{pages}{043015}
  (\bibinfo{year}{2010}).

\bibitem[{\citenamefont{Zhao and Babikov}(2006)}]{Zhao2006}
\bibinfo{author}{\bibfnamefont{M.}~\bibnamefont{Zhao}} \bibnamefont{and}
  \bibinfo{author}{\bibfnamefont{D.}~\bibnamefont{Babikov}},
  \bibinfo{journal}{J. Chem. Phys.} \textbf{\bibinfo{volume}{125}},
  \bibinfo{pages}{024105} (\bibinfo{year}{2006}).

\bibitem[{\citenamefont{Zhu and Rabitz}(1998)}]{Zhu1998b}
\bibinfo{author}{\bibfnamefont{W.}~\bibnamefont{Zhu}} \bibnamefont{and}
  \bibinfo{author}{\bibfnamefont{H.}~\bibnamefont{Rabitz}},
  \bibinfo{journal}{J. Chem. Phys.} \textbf{\bibinfo{volume}{109}},
  \bibinfo{pages}{385} (\bibinfo{year}{1998}).

\bibitem[{\citenamefont{Ohtsuki et~al.}(2007)\citenamefont{Ohtsuki, Teranishi,
  Saalfrank, Turinici, and Rabitz}}]{Ohtsuki2007}
\bibinfo{author}{\bibfnamefont{Y.}~\bibnamefont{Ohtsuki}},
  \bibinfo{author}{\bibfnamefont{Y.}~\bibnamefont{Teranishi}},
  \bibinfo{author}{\bibfnamefont{P.}~\bibnamefont{Saalfrank}},
  \bibinfo{author}{\bibfnamefont{G.}~\bibnamefont{Turinici}}, \bibnamefont{and}
  \bibinfo{author}{\bibfnamefont{H.}~\bibnamefont{Rabitz}},
  \bibinfo{journal}{Phys. Rev. A} \textbf{\bibinfo{volume}{75}},
  \bibinfo{pages}{033407} (\bibinfo{year}{2007}).

\bibitem[{\citenamefont{Takeushi}(2000)}]{Takeushi2000}
\bibinfo{author}{\bibfnamefont{S.}~\bibnamefont{Takeushi}},
  \bibinfo{journal}{Phys. Rev. A} \textbf{\bibinfo{volume}{62}},
  \bibinfo{pages}{032301} (\bibinfo{year}{2000}).

\bibitem[{\citenamefont{Dorai et~al.}(2000)\citenamefont{Dorai, Arvind, and
  Kumar}}]{Dorai2000}
\bibinfo{author}{\bibfnamefont{K.}~\bibnamefont{Dorai}},
  \bibinfo{author}{\bibnamefont{Arvind}}, \bibnamefont{and}
  \bibinfo{author}{\bibfnamefont{A.}~\bibnamefont{Kumar}},
  \bibinfo{journal}{Phys. Rev. A} \textbf{\bibinfo{volume}{61}},
  \bibinfo{pages}{042306} (\bibinfo{year}{2000}).

\bibitem[{\citenamefont{Ju et~al.}(2010)\citenamefont{Ju, Zhu, Peng, Chong,
  Zhou, and Du}}]{Ju2010}
\bibinfo{author}{\bibfnamefont{C.}~\bibnamefont{Ju}},
  \bibinfo{author}{\bibfnamefont{J.}~\bibnamefont{Zhu}},
  \bibinfo{author}{\bibfnamefont{X.}~\bibnamefont{Peng}},
  \bibinfo{author}{\bibfnamefont{B.}~\bibnamefont{Chong}},
  \bibinfo{author}{\bibfnamefont{X.}~\bibnamefont{Zhou}}, \bibnamefont{and}
  \bibinfo{author}{\bibfnamefont{J.}~\bibnamefont{Du}}, \bibinfo{journal}{Phys.
  Rev. A} \textbf{\bibinfo{volume}{81}}, \bibinfo{pages}{012322}
  (\bibinfo{year}{2010}).

\end{thebibliography}

\end{document}